\documentclass[twoside,11pt]{article}

\usepackage{jmlr2e}

\usepackage{blindtext}
\usepackage{caption}
\usepackage{amsmath,amssymb,amsfonts}
\usepackage[makeroom]{cancel}
\usepackage{graphicx} 
\usepackage{dsfont}
\usepackage[dvipsnames]{xcolor}
\DeclareMathOperator*{\argmin}{argmin}
\usepackage{bbm}
\usepackage{enumerate}
\usepackage{booktabs}
\usepackage{longtable}
\usepackage{bm}
\usepackage{float}

\usepackage{lastpage}

\ShortHeadings{Targeted tuning of random forests for quantile estimation and prediction intervals }{M. Berkowitz et al.}
\firstpageno{1}

\begin{document}

\title{Targeted tuning of random forests for quantile estimation and prediction intervals}

\author{\name Matthew Berkowitz \email
        mberkowi@sfu.ca \\
       \addr Statistics and Actuarial Science\\
       Simon Fraser University\\
       Burnaby, BC, Canada
       \AND
       \name Rachel MacKay Altman \email rachelm@sfu.ca \\
       \addr Statistics and Actuarial Science\\
       Simon Fraser University\\
       Burnaby, BC, Canada
       \AND
       \name Thomas M. Loughin \email tloughin@sfu.ca \\
       \addr Statistics and Actuarial Science\\
       Simon Fraser University\\
       Burnaby, BC, Canada}

\editor{TBD}

\maketitle

\begin{abstract}
We present a novel tuning procedure for random forests (RFs) that improves the accuracy of estimated quantiles and produces valid, relatively narrow prediction intervals. While RFs are typically used to estimate mean responses (conditional on covariates), they can also be used to estimate quantiles by estimating the full distribution of the response. However, standard approaches for building RFs often result in excessively biased quantile estimates. To reduce this bias, our proposed tuning procedure minimizes ``quantile coverage loss" (QCL), which we define as the estimated bias of the marginal quantile coverage probability estimate based on the out-of-bag sample. We adapt QCL tuning to handle censored data and demonstrate its use with random survival forests. We show that QCL tuning results in quantile estimates with more accurate coverage probabilities than those achieved using default parameter values or traditional tuning (using MSPE for uncensored data and C-index for censored data), while also reducing the estimated MSE of these coverage probabilities. We discuss how the superior performance of QCL tuning is linked to its alignment with the estimation goal. Finally, we explore the validity and width of prediction intervals created using this method.
\end{abstract}

\begin{keywords}
  random forest, prediction interval, quantile estimation, tuning, random survival forest
\end{keywords}

\section{Introduction}
\label{sec1}

Random forests (RFs; \citealp{breiman2001}) for regression are a class of ensemble methods where regression trees are grown on $B$ bootstrap resamples or subsamples of a training dataset for some large $B$. Each tree is constructed by recursively splitting data into smaller and smaller subsets called ``nodes". At each tree node, a subset of the $p$ covariates are randomly selected as candidates upon which to base a split, and the optimal split location is chosen to optimize some loss function, usually squared-error loss. The splitting process ends in ``terminal nodes" representing partitions of the covariate space. In a standard regression tree, the sample mean of all responses within each terminal node is computed. Then, for a given covariate value, the RF provides an estimate of the conditional mean of the response by averaging the sample means from all trees' terminal nodes that are associated with this value. An important feature of RFs is that the sampling process for each tree excludes some of the training data from its construction. These excluded observations are referred to as ``out of bag'' (OOB) data and provide RFs with a built-in validation set for error estimation and tuning. 

The vast majority of research on RFs has focused on this original formulation, where only sample means are computed within terminal nodes and averaged across trees. But \cite{meinshausen} recognized that other information could be extracted from the responses in the terminal nodes, specifically, the empirical distribution function (EDF). He demonstrated that, under certain restrictive regularity conditions, the average of these EDFs across trees of the forest provides a consistent estimate of the entire conditional distribution function of the response, given the covariate value. He proposed using this estimated conditional distribution function (ECDF) for quantile estimation and referred to this method as a quantile regression forest (QRF). \cite{athey} extended these ideas to a class of RFs where the tree construction is based on loss functions other than squared error, resulting again in an ECDF that is a consistent estimate of the true conditional distribution under regularity conditions.  

RFs have been adapted to numerous other data structures, such as right-censored data, which are common when responses are times until an event occurs. The most influential RF adaptation for right-censored data is the random survival forest (RSF) \citep{ishwaran2008}. The algorithm for building RSFs is similar to that for standard RFs except that the loss function used to determine splits must be one that is appropriate for right-censored data. The log-rank statistic is used most commonly, although others have been suggested (e.g., \citealp{moradian}). Similar to the approach of \cite{meinshausen}, the observations in each terminal node are used to form a Nelson-Aalen estimate of the cumulative hazard function (CHF) or Kaplan-Meier estimate of the survival function within the node. The estimates associated with a given covariate value are then averaged across trees to obtain an ensemble CHF or survival function estimate. Estimated survival probabilities or quantiles can easily be obtained from either estimate.

A commonly cited advantage of RFs is that they often produce adequate estimates of the mean response even without tuning \citep{friedman}. More recent studies question this conventional wisdom \citep{ishwaran2011, mentch2020}. Tuning in RFs focuses on two main quantities: the number of covariates that are randomly selected into the pool of candidates when a node is under consideration for splitting, often called \texttt{mtry}, and the size of trees grown on each sample. Tree size may be controlled in several ways; the one most commonly used in the literature and in popular software implementations, called \texttt{nodesize}, governs how large a node must be to allow further splits to take place. The default values for the standard RF regression setting, \texttt{mtry} $=p/3$ and \texttt{nodesize} $=5$, are largely based on limited simulation results dating back to \cite{breiman2001}. Our work shows that default values and conventional tuning approaches (such as tuning to minimize the mean square prediction error, MSPE) may lead to biased quantile estimates and invalid prediction intervals. 

The literature on tuning RFs is sparse. In a narrative review, \cite{biauscornet} found a paucity of papers that have investigated tuning RFs and essentially no theoretical support for default values of \texttt{mtry} and \texttt{nodesize}. Other theoretical results focus on large-sample properties and do not directly address the practical implications of tuning for finite-sample performance \citep{breiman2004, ishwaran2010, biau2010, biau2012, denil, wager, scornet2015, meinshausen, edcosaque}. \cite{scornet2017} noted that, although his prior consistency proofs \citep{scornet2015} of RFs are valid for any value of \texttt{mtry}, in practice (where the sample size is finite), \texttt{mtry} needs to be tuned. \cite{durouxscornet} demonstrated theoretically and empirically that tuning the subsampling rate and tree depth in random forests can improve their performance substantially. 

More recent finite-sample investigations by \cite{mentch2020} found, among other results, that high \texttt{mtry} values work better (in terms of minimizing the mean squared error of prediction) in high signal-to-noise ratio (SNR) settings and low \texttt{mtry} values work better in low SNR settings. Additionally, they found---and reaffirmed in a follow-up paper \citep{mentch2022}---benefits from the injection of noise variables in low SNR settings, explaining that these variables act as an implicit regularization mechanism. Furthermore, contrary to conventional RF tuning wisdom, researchers have found that tuning tree depth can also have benefits in lower SNR settings \citep{zhou, surjanovic}.

Whether or how to tune RFs to estimate quantities other than the mean is a question that has hardly been explored. \cite{ishwaran2011} studied the relationship among tree depth, variable selection, and tuning parameters on survival probability estimation in random survival forests (RSFs) with high-dimensional data. They found that using higher \texttt{mtry} values improves the chance of splitting on strong variables, improving the forest's ability to discover the signal in high SNR scenarios---foreshadowing the findings of \cite{mentch2020}.

However, our focus is on quantile estimation in RFs and related ensembles. Are conventional tuning methods or forests using default tuning parameters capable of producing quantile estimates with accurate coverage probabilities? And does the accuracy vary by the choice of quantile and by the data-generating mechanism?  

Many RF-focused papers have argued for the importance of aligning the \textit{splitting criterion} with the estimation goal or evaluation criterion (e.g., \citealp{schmid, moradian, athey}). In this paper, we argue that the loss function that we use for \textit{tuning} should also correspond to the estimation goal or evaluation criterion. Specifically, we study RF tuning when the goal is estimating quantiles with accurate coverage probabilities. Although biases in the lower- and upper-quantile estimates can sometimes offset each other so that a prediction interval still has the nominal coverage rate, accurate marginal coverage at each target quantile can be important for producing intervals that have accurate coverage and appropriate width. Accordingly, we propose tuning using the estimated marginal bias in estimated quantile coverage probability as the primary loss function. We show that tuning using this loss function substantially improves the accuracy of estimated quantiles in terms of their coverage probabilities compared to not tuning (i.e., using default tuning parameters) or tuning conventionally. We also show that our tuning approach can be used to produce valid prediction intervals that are competitive with other methods of producing RF-based prediction intervals.

We begin in Section~\ref{sec2} by briefly presenting some empirical results that demonstrate the inadequacy of untuned or conventionally tuned RFs for estimating quantiles. In Section~\ref{sec3}, we describe in detail our proposed tuning procedure---quantile coverage loss (QCL)---to reduce the bias of quantile coverage probabilities. In Section~\ref{sec4}, we discuss the design of our simulation study and the metrics we use to evaluate the performance of our tuned forests. In Section~\ref{sec5}, we discuss our results and make recommendations for practitioners that are tailored to specific goals. Moreover, we offer explanations for why our tuning procedure excels across diverse conditions and maintains a clear edge over no tuning or traditional tuning, even in cases where all methods struggle to produce quantile estimates with accurate coverage probability, and conclude by underscoring our main contributions, outlining our procedure's limitations, and providing avenues for future research opportunities.

\section{The Problem}
\label{sec2}

In Section~\ref{sec4}, we describe a simulation study comparing the estimated bias in coverage probabilities of quantile estimates obtained from RFs under 108 different simulation settings with normally distributed responses. For each setting, we tuned RFs in a variety of ways and used these tuned RFs to estimate the 0.1 quantiles corresponding to 1000 test observations. We estimated the coverage probability of the quantile estimates produced by each RF by computing the theoretical quantile coverage probability for each test observation's quantile estimate using the true distribution. We repeated this process over 10 training sets per simulation setting and estimated the marginal coverage probability bias as the average difference between the estimated coverage probability and target coverage probability (0.1) across all test observations.  

\begin{figure}[ht]
\centering
\includegraphics[width=1.0\textwidth]{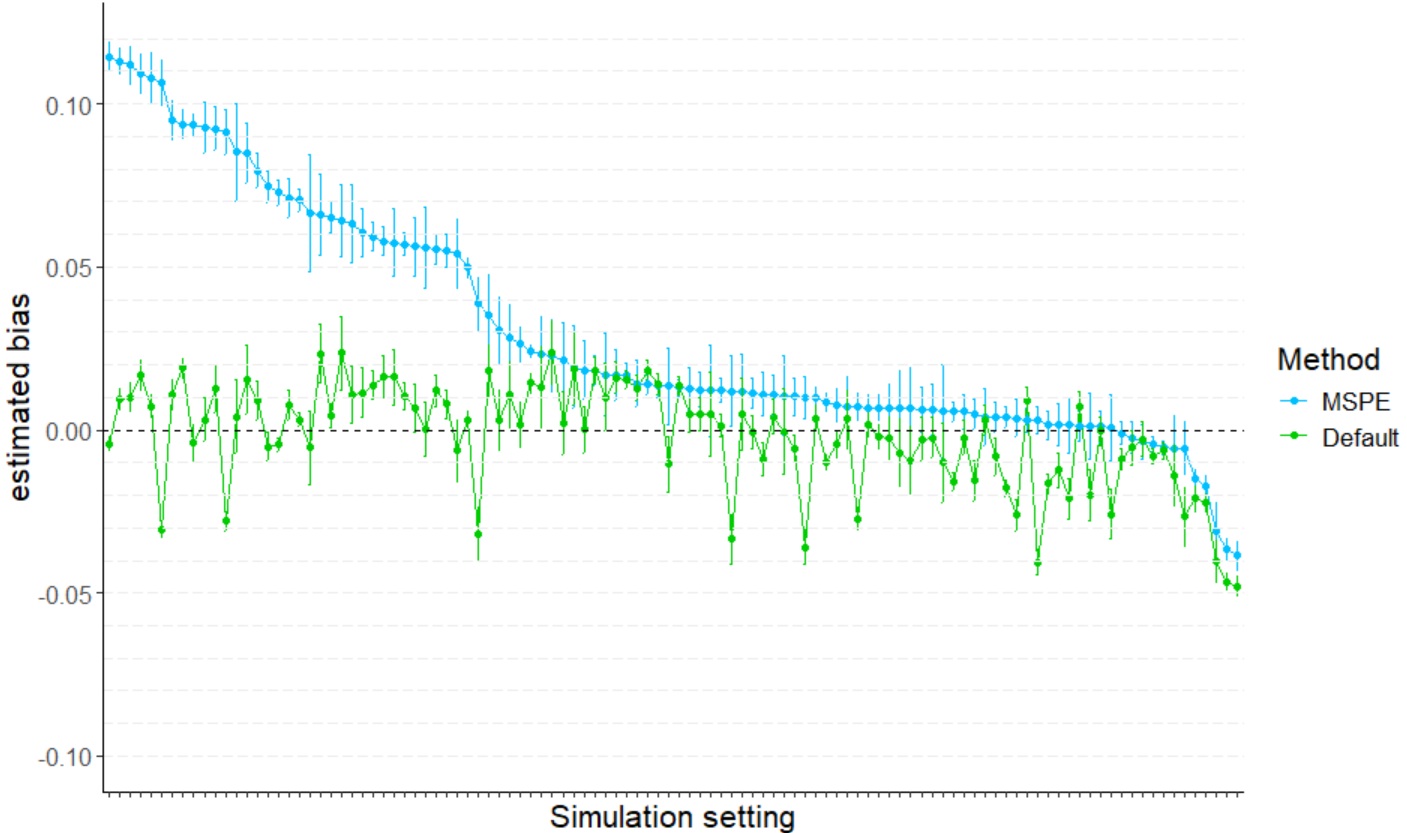}
\caption{Estimated marginal coverage probability bias for 0.1 quantile using MSPE tuning and default tuning parameter values}
\label{fig_problem}
\end{figure}

Figure~\ref{fig_problem} displays the estimated marginal coverage probability biases for each simulation setting when estimating the 0.1 quantile using no tuning and traditional MSPE tuning. These estimated biases are plotted in descending order based on MSPE tuning. The error bars on the plot represent 95\% confidence intervals for the mean bias within each simulated setting.

In the vast majority of simulation settings, the confidence intervals for the coverage bias exclude zero for both MSPE-tuned and untuned RFs. In other words, both methods produce quantile estimates with significantly biased coverage probabilities under many realistic settings (see Section~\ref{sec5} for details on the settings). 

Clearly, we need a better approach to RF-based quantile estimation.  

\section{QCL Tuning for Quantile Estimation}
\label{sec3}

In this section, we describe our RF tuning approach for estimating quantiles with accurate coverage probabilities.

We use the following notation. Let $T$ denote the numerical response of a randomly chosen individual, and let $\mathbf{X} = (X_1, \ldots, X_p)$ denote a $p$-dimensional vector of covariates. We observe a random sample from the joint distribution of $(T,\mathbf{X})$,$(t_i, \mathbf{x}_i),\, i=1,\ldots,N$, where the first $n$ pairs form the training set, and the remaining $N-n$ pairs form the test set. 

One special case of interest occurs when $T$ represents a time to an event. In this case, responses may sometimes be subject to right-censoring, where the event time is not observed but is known to have occurred after some observed ``censoring time''. Estimating quantiles using such censored data is particularly important, for example, in oncology \citep{hong, yazdani}. Let $C$ denote the censoring time for a randomly chosen individual. The observable response time is then defined by $Y = \min(T, C)$. Let $\Delta = \mathbbm{1}(T \leq C)$ indicate when an observable time is an event time rather than a censoring time. Consequently, the observed data are given by $(y_i, \delta_i, \mathbf{x}_i),\, i=1,\ldots,N$.

In all cases, the conditional distribution function of $T$ given covariates $\mathbf{X} = \mathbf{x}$ is denoted by $F_T(t \mid \mathbf{x})$. The conditional quantile function for the $\tau$ quantile, given the covariates, is denoted by $q_{\tau}(\mathbf{x}) = F_T^{-1}(\tau \mid \mathbf{x})$, for $0<\tau<1$. When no observations are censored, we use RFs to estimate $F_T$ as per \cite{meinshausen}; when some observations are censored, we estimate $F_T$ via the estimated survival function that is produced using the RSF method of \cite{ishwaran2008}. However, our arguments are not limited to these specific methods; our approach applies equally to distribution function estimates produced by other techniques.

\subsection{Estimating the marginal coverage probability of a quantile estimate}
\label{sec3.1}

Our goal is to tune RFs to produce quantile estimates whose coverage probabilities are as accurate as possible---ideally, for any given value of the covariates. As \cite{zhang} point out, this task is very challenging in general. Instead, we strive to produce quantile estimates with accurate {\em marginal} coverage, averaged across the distribution of $\mathbf{X}$, as other authors have done in the context of prediction intervals (e.g., \citealp{zhang}).

Specifically, let $\hat{q}_{\tau}(\mathbf{x})=\hat{F}_T^{-1} (\tau|\mathbf{x})=\textrm{inf} \{t: \hat{F}_T(t | \mathbf{x}) \geq \tau \}$ be the RF estimate of the $\tau$ quantile at $\mathbf{x}$, where $\hat{F}_T(\cdot|\mathbf{x})$ is the RF estimate of the CDF of $T$ for covariates $\mathbf{x}$ (the OOB estimate in the case where $\mathbf{x}$ is from the training data). Note that $\hat{F}(\cdot)$ and $\hat{q}_{\tau}(\cdot)$ depend on tuning parameters, but we suppress this notation for convenience. The marginal coverage probability, $\tilde{\tau}$, of the estimate for the $\tau$ quantile is
\begin{align}
\label{eq1}
    \tilde{\tau} = \int_{\cal \mathbf{X}} F_T (\hat{q}_{\tau}(\mathbf{x})|\mathbf{x})dF_\mathbf{X}(\mathbf{x}),
\end{align}
where $F_\mathbf{X}(\mathbf{x})$ is the marginal distribution of $\mathbf{X}$. When the responses are fully observed, we use
\begin{align}
\label{eq2}
    \hat{\tilde{\tau}} =  \frac{1}{n}\sum_{i=1}^n \hat{F}_{i,T}(\hat{q}_{\tau}(\mathbf{x}_i)|\mathbf{x}_i)
\end{align}
as the estimate of $\tilde{\tau}$ and 
\begin{align}
\label{eq3}
    \hat{F}_{i,T}(\hat{q}_{\tau}(\mathbf{x}_i)|\mathbf{x}_i) = \mathbbm{1}(t_i \leq \hat{q}_{\tau}(\mathbf{x}_i)), i=1,\ldots,n,
\end{align}  
as the estimate of the CDF for observation $i$ in the training set. We discuss the estimation of these quantities in the censored-data setting in Section~\ref{sec3.3}.

\subsection{Quantile coverage loss}
\label{sec3.2}

Consider training a RF where the tuning parameters are specified in a vector $\bm \theta$. Since our goal is to select $\bm \theta$ to minimize the absolute bias of the marginal coverage probability, we define population quantile coverage loss (QCL) as 
\begin{align}
\label{eq4}
    \mathcal{L}_\tau(\bm{\theta}) = \left| \tilde{\tau}- \tau \right|,
\end{align} and we estimate it using 
\begin{align}
\label{eq5}
    \hat{\mathcal{L}}_\tau(\bm{\theta}) =  \left| \hat{\tilde{\tau}}- \tau \right|.
\end{align}
Both $\tilde{\tau}$ and $\hat{\tilde{\tau}}$ depend on $\bm{\theta}$, but we have suppressed this dependence for ease of notation.

When tuning RFs to estimate quantiles for a given $\tau$, the goal is to identify the tuning parameter combination that minimizes this loss function. We refer to this process as QCL tuning. Importantly, the optimal tuning parameters may vary depending on the probability level $\tau$ at which quantiles are estimated. Therefore, we conduct the tuning process for each $\tau$ at which quantiles are sought. In particular, for prediction intervals, we run the tuning procedure twice, once for each endpoint. 

For example, suppose that we wish to simultaneously estimate the 0.1 and 0.9 quantiles of some distribution conditional on $\mathbf{X}$. If the tuning algorithm is one such as grid search or random selection where all values of $\bm{\theta}$ are known prior to fitting the forests---say, $\bm{\theta}_1,\ldots,\bm{\theta}_K$---then a single RF can be fit using each selected $\bm{\theta}_k$. From forest $k$, we can use the OOB estimate of $F_T(t|\mathbf{x}_i)$ to find both $\hat{q}_{0.1}(\mathbf{x}_i)$ and $\hat{q}_{0.9}(\mathbf{x}_i)$ for $i=1,\ldots,n$, which can then be used to compute $\hat{\mathcal{L}}_{0.1}(\bm{\theta_k})$ and $\hat{\mathcal{L}}_{0.9}(\bm{\theta_k})$. The final tuning parameters can be found separately for each $\tau$ as $$\bm{\theta}_{\mbox{\scriptsize opt},\tau}=\argmin_k(\hat{\mathcal{L}}_{\tau}(\bm{\theta_k})).$$
On the other hand, if a sequential algorithm such as model-based optimization is used (e.g., \citealp{hutter}), then the entire algorithm must be rerun for each value of $\tau$. 

To construct $100(1-\alpha)\%$ prediction intervals with equal tail probabilities using QCL tuning, we fit separate RFs using each $\bm{\theta}_k,\,k=1,\ldots,K$, for $\tau=\alpha/2$ and $\tau=1-\alpha/2$ as described above. We then consider all $K^2$ pairings from these two sets of RFs and select the combination that produces the narrowest interval while maintaining a coverage probability of at least $100(1-\alpha)\%$. This strategy prioritizes meeting a minimum coverage probability requirement over reducing coverage probability bias in the two endpoints separately, and it has produced intervals with better coverage and width properties in preliminary testing.

A similar amendment could be applied when an individual quantile is estimated for use as the endpoint of a one-sided prediction interval, which may be of particular interest with time-to-event outcomes. Rather than minimizing the absolute bias as in \eqref{eq4} and \eqref{eq5}, one could instead minimize bias subject to its being nonnegative or nonpositive, depending on whether an upper or lower endpoint, respectively, is sought (to increase the chance that the procedure will produce a valid prediction interval). 

\subsection{Estimating \eqref{eq1} with censored data}
\label{sec3.3}

When the response $T$ is an event time subject to right censoring, computing \eqref{eq3} is not always possible. Specifically, when $y_i$ is a censored time, the exact event time $t_i$ is an unknown value such that $t_i > y_i$, and we may not be able to evaluate $\mathbbm{1}(t_i \leq \hat{q}_{\tau}(\mathbf{x}_i))$. In addition, with RSFs and other RFs for censored data, the ECDF may plateau at a probability value, say $\tau^*$, well below 1, that can vary depending on $\mathbf{x}$. (For a given terminal node in a given tree, the K-M estimate of the survival function is defined only up to the largest event time in the node. But in the RSF package and other survival forest implementations, the estimated survival function for any time between the largest event time in the node and the largest event time in the training set is defined as the value of the function at the largest event time in the node.) Consequently, quantile estimates for any value of $\tau>\tau^*$ are not defined. Therefore, we describe two possible approaches for estimating $\tilde{\tau}$ from RSFs and related ensembles. The first approach, which we call QCL-C, uses a different estimate of $F_T(\hat{q}_{\tau}(\mathbf{x})|\mathbf{x})$ in \eqref{eq2} for observations where the indicator in \eqref{eq3} cannot be evaluated. The second approach, called QCL-IPCW, uses inverse-probability-of-censoring weights.

\paragraph{QCL-C.} Let $\hat{q}_{\tau^*}(\mathbf{x}_i)$ be the maximum quantile estimate that is defined based on the OOB conditional distribution function estimate for observation $i$. When $\tau > \tau^*$, $\hat{q}_{\tau}(\mathbf{x}_i)$ is a hypothetical estimate of the quantile we seek; we need make no assumptions about it other than $\hat{q}_{\tau}(\mathbf{x}_i) \ge \hat{q}_{\tau^*}(\mathbf{x}_i)$. With censored data, to estimate \eqref{eq1}, we need to consider four cases:

\begin{enumerate}[(I)]
    \item $T_i = y_i$ and $\hat{q}_{\tau}(\mathbf{x}_i) \le  \hat{q}_{\tau^*}(\mathbf{x}_i)$
    \item $T_i > y_i$ and $\hat{q}_{\tau}(\mathbf{x}_i) \le  \hat{q}_{\tau^*}(\mathbf{x}_i)$
    \item $T_i = y_i$ and $\hat{q}_{\tau}(\mathbf{x}_i) > \hat{q}_{\tau^*}(\mathbf{x}_i)$
    \item $T_i > y_i$ and $\hat{q}_{\tau}(\mathbf{x}_i) > \hat{q}_{\tau^*}(\mathbf{x}_i)$.
\end{enumerate}

Case I is the usual situation that arises in the absence of censoring, i.e., we can evaluate \eqref{eq3} directly in this case. The other three cases present challenges. Cases II and IV refer to observations whose event times are censored, and, hence, we cannot definitively establish the ordering of $t_i$ and $\hat{q}_{\tau}(\mathbf{x}_i)$. Cases III and IV deal with the situation where the desired quantile is not defined. 

For these cases, we decompose the integrand probability in \eqref{eq1} as follows:
\begin{align}
    & F_T ( \hat{q}_{\tau}(\mathbf{x}_i)|\mathbf{x}_i) \nonumber \\
    &=P(T_i \leq \hat{q}_{\tau}(\mathbf{x}_i)|T_i\leq C_i, \mathbf{x}_i) P(T_i \leq C_i|\mathbf{x}_i) + 
    P(T_i \leq \hat{q}_{\tau}(\mathbf{x}_i)|T_i>C_i, \mathbf{x}_i) P(T_i>C_i|\mathbf{x}_i)\nonumber \\
    &=P(T_i \leq \hat{q}_{\tau}(\mathbf{x}_i)|T_i=Y_i, \mathbf{x}_i) P(\Delta_i=1|\mathbf{x}_i) + 
    P(T_i \leq \hat{q}_{\tau}(\mathbf{x}_i)|T_i>Y_i, \mathbf{x}_i) P(\Delta_i=0|\mathbf{x}_i).
\label{eq6}
\end{align}
Plugging in sample quantities, we estimate $P(\Delta_i=1|\mathbf{x}_i)$ with $\delta_i$ and $P(\Delta_i=0|\mathbf{x}_i)$ with $1-\delta_i$.
To estimate $P(T_i \leq \hat{q}_{\tau}(\mathbf{x}_i)|T_i=Y_i, \mathbf{x}_i)$ and $P(T_i \leq \hat{q}_{\tau}(\mathbf{x}_i)|T_i>Y_i, \mathbf{x}_i)$, we need to consider whether $\hat{q}_{\tau}(\mathbf{x}_i)$ is defined for the specified $\tau$. 

\begin{enumerate}[\indent {}]
    \item \textbf{Case II}. When the observed time is censored at a time $y_i \ge \hat{q}_{\tau}(\mathbf{x}_i)$, then $t_i$ must be greater than $\hat{q}_{\tau}(\mathbf{x}_i)$, too. Thus, we can evaluate the indicator in \eqref{eq3}. However, when $y_i < \hat{q}_{\tau}(\mathbf{x}_i)$, then we do not know the ordering of $t_i$ and  $\hat{q}_{\tau}(\mathbf{x}_i)$. In this case, we estimate $P(T_i \leq \hat{q}_\tau(\mathbf{x}_i) \mid T_i > Y_i, \mathbf{x}_i)$ in \eqref{eq6} as 
        \begin{align}
        \label{eq7}
            \hat{P}(T_i \leq \hat{q}_\tau(\mathbf{x}_i) \mid T_i > y_i, \mathbf{x}_i) &= 1 - \hat{P}(T_i > \hat{q}_\tau(\mathbf{x}_i) \mid T_i > y_i, \mathbf{x}_i) \nonumber \\ \nonumber 
            &= 1 - \frac{\hat{P}(T_i > \hat{q}_\tau(\mathbf{x}_i)|\mathbf{x}_i)}{\hat{P}(T_i > y_i|\mathbf{x}_i)} \\ 
            &= 1 - \frac{1 - \tau}{1-\hat{F}_T  (y_i|\mathbf{x}_i)}.
        \end{align}
        In this case, because $\hat{P}(\Delta_i=1|\mathbf{x}_i)=0$, the estimate of \eqref{eq6} reduces to \eqref{eq7}. 
        
        \item \textbf{Case III}. In RSF and related ensembles, the estimated conditional distribution function is defined only up to the largest observed event time. Hence, if $y_i$ is an event time, it must be no larger than the largest estimable quantile $\hat{q}_{\tau^*}(\mathbf{x}_i)$.  Therefore, if $\hat{q}_{\tau}(\mathbf{x}_i) > \hat{q}_{\tau^*}(\mathbf{x}_i)$, then we estimate $P(T_i \leq \hat{q}_\tau(\mathbf{x}_i) \mid T_i = Y_i, \mathbf{x}_i)$ in \eqref{eq6} as $\hat{P}(T_i \leq \hat{q}_{\tau}(x_i) \mid T_i=y_i, \mathbf{x}_i)=1$. Because $\hat{P}(\Delta_i=1|\mathbf{x}_i)=1$, the estimate of \eqref{eq6} reduces to 1 as well.

        \item \textbf{Case IV}. If $y_i \le \hat{q}_{\tau^*}(\mathbf{x}_i)$, then $y_i \le \hat{q}_\tau(\mathbf{x}_i)$ because $\hat{q}_{\tau^*}(\mathbf{x}_i)$ is always less than or equal to $ \hat{q}_{\tau}(x_i)$. So we define
        \begin{align}
        \label{eq8}
            \hat{P}(T_i \leq \hat{q}_{\tau}(x_i) \mid T_i > y_i, \mathbf{x}_i ) = 1 - \frac{1-\tau}{1-\hat{F}_T(y_i\mid x_i)}
        \end{align}
        as in \eqref{eq7}. In this case, because $\hat{P}(\Delta_i=1|\mathbf{x}_i)=0$, the estimate of \eqref{eq6} reduces to \eqref{eq8}. 

        On the other hand, if $\hat{q}_{\tau^*}(\mathbf{x}_i)< y_i$, then we cannot know the ordering of $y_i$ and $\hat{q}_\tau(\mathbf{x}_i)$, and we cannot estimate $F_T(y_i\mid \mathbf{x}_i)$. But using \eqref{eq7}, we do know that 
        \begin{align} 
        \label{eq9}
            \hat{P}(T_i \leq \hat{q}_{\tau}(\mathbf{x}_i) \mid T_i > y_i, \mathbf{x}_i) 
            &\leq 1-\frac{1-\tau}{1-\hat{F}_T(\hat{q}_{\tau^*}(\mathbf{x}_i) \mid \mathbf{x}_i)} \nonumber \\
            &= 1-\frac{1-\tau}{1-\tau^*} \nonumber \\
            &=\frac{\tau - \tau^*}{1 - \tau^*}.
        \end{align}

        We use the upper bound \eqref{eq9}
        as our estimate of $P(T_i \leq \hat{q}_{\tau}(\mathbf{x}_i) \mid T_i > y_i, \mathbf{x}_i)$ in \eqref{eq6}. Then, because $\hat{P}(\Delta_i=1|\mathbf{x}_i)=0$, the estimate of \eqref{eq6} reduces to \eqref{eq9}. 
    
\end{enumerate}
We summarize these four cases in Table~\ref{table_qclccases}.

\begin{table}[ht]
\begin{tabular}{@{}llll@{}}
\toprule
\textbf{Case} & \textbf{$\hat{q}_{\tau}(\mathbf{x}_i)$ defined} & \textbf{$\delta_i$} & \textbf{$\hat{F}_{i,T}(\hat{q}_{\tau}(\mathbf{x}_i)|\mathbf{x}_i)$} \\ \midrule
I   & Yes & 1 & $\displaystyle \mathbbm{1}\Bigl(t_i \leq \hat{q}_\tau(\mathbf{x}_i)\Bigr)$ \\ \midrule
II  & Yes & 0 & \begin{tabular}[c]{@{}l@{}}$\displaystyle \mathbbm{1}\Bigl(t_i \leq \hat{q}_\tau(\mathbf{x}_i)\Bigr)$, if $y_i \ge \hat{q}_{\tau}(\mathbf{x}_i)$\\[4pt]$\displaystyle 1 - \frac{1 - \tau}{1-\hat{F}_T  (y_i|\mathbf{x}_i)}$, if $y_i < \hat{q}_{\tau}(\mathbf{x}_i)$\end{tabular} \\ \midrule
III & No & 1 & 1 \\ \midrule
IV  & No & 0 & \begin{tabular}[c]{@{}l@{}}$\displaystyle \frac{\tau - \tau^*}{1 - \tau^*}$, if $y_i > \hat{q}_{\tau^*}(\mathbf{x}_i)$\\[4pt]$\displaystyle 1 - \frac{1 - \tau}{1-\hat{F}_T  (y_i|\mathbf{x}_i)}$, if $y_i \leq \hat{q}_{\tau^*}(\mathbf{x}_i)$\end{tabular} \\ \bottomrule
\end{tabular}
\captionsetup{justification=raggedright,singlelinecheck=false}
\caption{QCL-C's four cases}
\label{table_qclccases}
\end{table}

\paragraph{QCL-IPCW.}
As an alternative to QCL-C, inverse probability of censoring weights (IPCW, \citealp{lawlessyuan}) can be used to estimate quantile coverage probability. In particular, to estimate \eqref{eq1}, we theoretically could use the estimate
\begin{align}
\label{eq10}
    \frac{1}{n}\sum_{i=1}^n \frac{\Psi_i \mathbbm{1}(t_i \leq \hat{q}_\tau(\mathbf{x}_i))}{\hat{\alpha}_i},
\end{align}
where
\[
\Psi_i =
  \begin{cases}
    1, & \mbox{if  $\hat{q}_\tau(\mathbf{x}_i)\le \hat{q}_{\tau^*}(\mathbf{x}_i)$ and either $\delta_i=1$ or $\delta_i=0$ and $y_i> \hat{q}_\tau(\mathbf{x}_i)$} \\
    0, & \mbox{otherwise}
  \end{cases}
\]
and $\hat{\alpha}_i$ is an estimate of $\alpha_i=P(\Psi_i=1 \mid T_i=t_i)$, assumed common across covariate values. However, an unbiased estimator of this probability is not readily available. We therefore approximate $P(\Psi_i=1 \mid T_i=t_i)$ with $P(C_i>t_i)$ and estimate the latter via the observed censoring times. Our estimate of $\tilde{\tau}$ is then  
\begin{align}
\label{eq11}
    \hat{\tilde{\tau}}_{w} = \frac{1}{n}\sum_{i=1}^n \frac{\Psi_i \mathbbm{1}(t_i \leq \hat{q}_\tau(\mathbf{x}_i))}{\hat{P}(C_i > t_i)}.
\end{align}
Note that $\mathbbm{1}(t_i \leq \hat{q}_\tau(\mathbf{x}_i))$ can't be evaluated if the $i^{th}$ observation is censored and $y_i \le \hat{q}_\tau(\mathbf{x}_i)$. But in this case, $\Psi_i=0$, so the entire numerator is also zero.

To estimate $P(C_i > t_i)$, we use the Kaplan-Meier estimator of censoring times (i.e., we assume that the censoring times are iid). But $\alpha_i$ and $P(C_i>t_i)$ could be modelled conditional on the covariates, in which case other estimators of $P(C_i>t_i)$ in \eqref{eq11} could be considered (e.g., see \citealp{libradic}). All estimators (including the Kaplan-Meier estimator) require assumptions about the distribution of the censoring times. If these assumptions don't hold, the estimator of $\tilde{\tau}$ could be biased. In constrast, the QCL-C estimator doesn't require such assumptions.

\section{Design of Simulation Studies}
\label{sec4}

In this section, we describe our simulation study, which we designed to compare the properties of quantile estimates obtained using QCL-tuned, conventionally tuned, and forests using default tuning paramters (see Section~\ref{sec4.3}). We also define the error metrics we used to evaluate performance. The comparisons were made in the context of four different problems: estimating quantiles and forming prediction intervals using fully observed data and RFs, and  estimating quantiles and forming prediction intervals using censored data and RSFs. 

To estimate quantiles with fully observed data, we compared four methods: RFs tuned using QCL as per Section~\ref{sec3.2}, RFs tuned with MSPE (the standard loss function), RFs with default tuning parameters, and generalized random forests (GRFs) with their default parameter values \citep{athey}. We included GRFs because they use a locally adaptive splitting rule that the authors applied to quantile estimation as a test case, demonstrating superior performance compared to the usual squared-error-loss splitting criterion in the settings they considered.

For prediction intervals, we compared QCL-tuned intervals (as described in Section~\ref{sec3.2}) to untuned QRF intervals, which use the ECDF produced by a single RF built using default tuning parameters to estimate lower and upper quantiles \citep{meinshausen}. We also computed two residual-based intervals, which we call Res-OOB and Res-SC. 

Res-OOB assumes that the errors are iid and uses the OOB estimate of their common distribution \citep{zhang}. As a result, the intervals' widths are identical for all test observations (because they are based on the empirical quantiles of the OOB estimate of the error distribution). Res-OOB intervals are constructed using the empirical quantiles of the RF OOB prediction errors, $r_i=t_i-\hat{t}_{(i)}$, $i=1,\ldots,n$, where $\hat{t}_{(i)}$ is the OOB prediction (typically based on the mean estimate) for observation $i$. A $100(1-\alpha)\%$ prediction interval for a response, $T$, based on its predicted value, $\hat{t}$, is constructed as
\begin{center}
    $(\hat{t} - R_{\alpha/2}, \hat{t} + R_{1-\alpha/2})$,
\end{center}
where $R_{\tau}$ is the $\tau$ quantile of the EDF based on $r_1,\ldots,r_n$. As \cite{zhang} recommend when the distribution of errors is symmetric, we use the slightly modified interval $\hat{t} \pm \tilde{R}_{1-\alpha}$, where $\tilde{R}_{\tau}$ is the $\tau$ quantile of the EDF based on $|r_1|,
\ldots,|r_n|$. We implemented this latter version for our simulation study, which used homoscedastic normal errors.

Res-SC intervals use split conformal inference \citep{lei} and are constructed by first splitting the dataset into two equal halves (a training and a test set) and then fitting a RF to the training set. For each observation in the test set, $t_i$, the predicted value $\hat{t}_i$ is obtained and the absolute residual, $d_i= |t_i - \hat{t}_i|$, is computed, for $i=1,\ldots,n/2$. A $100(1-\alpha)\%$ prediction interval for a response, $T$, based on its predicted value using the $n/2$ observations in the training data, $\hat{t}$, is then constructed as
\begin{center}
    ($\hat{t}-D_{\alpha/2}, \hat{t}+D_{\alpha/2}$),
\end{center}
where $D$ is the $(1-\alpha)$ quantile of the EDF based on $d_1, \dots, d_{n/2}$.

In the censored-data setting, we compared four methods of estimating quantiles and forming prediction intervals using RSFs: tuning with each of QCL-C and QCL-IPCW as described in Section~\ref{sec3.3}, standard tuning with the concordance index (C-index), and using default tuning parameters. For prediction intervals, we compared QCL-C- and QCL-IPCW-tuned intervals to intervals produced by a single RSF built using default tuning parameters to estimate lower and upper quantiles.

The computational workflow we used is as follows, with details on each step given in subsequent sections:

\begin{enumerate}[(i)]
\item Generate training datasets under settings specified by combinations of several factors detailed in Section \ref{sec4.1}. For each training set, generate a companion test set of size 1000. There were 108 settings with uncensored data and 96 settings with censored data. For the censored setting, we generate only failure times in the test set. For each setting, we generated 10 training-test dataset pairs.  
\item On each training dataset:

\begin{enumerate}[(a)]
\item Fit RFs or RSFs using a grid of $\theta = (\texttt{mtry}, \texttt{nodesize})$ values. The grid is formed from every possible \texttt{mtry} value for the given setting and \texttt{nodesize} values $\{1,5,10,25,40\}$ for uncensored data and $\{3,8,15,30\}$ for censored data. (Note: we omitted $\texttt{nodesize}=1$ from the grid in the censored setting because it led to highly variable estimates of the survival function.) For uncensored data, we also fit GRFs using the default tuning parameter values.

\item Using these fitted forests, estimate the $\tau \in \{0.1,0.5,0.9\}$ quantiles for each observation in the training set, and estimate the marginal coverage probability using \eqref{eq2}.

\item For each $\tau$, tune using QCL as in \eqref{eq3}, and tune using MSPE (uncensored data) or C-index (censored data). In other words, we determine which value of $\theta$ minimizes each method's OOB loss function estimate, and select the forest associated with this optimal $\theta$. This process is conducted separately for each $\tau$ in the case of QCL tuning but just once per dataset otherwise.

\item Using the companion test set, compute error metrics (defined in Section~\ref{sec4.2}) for each forest's estimate of each quantile. Also identify the value of $\mathbf{\theta}$ whose forest has the smallest value of the error metric on the test dataset. Refer to that version of the forest as the ``Oracle'' for that dataset.
\end{enumerate}

\item For each $\tau$, summarize the error metrics from all data sets for all methods. For uncensored data, there are 1080 separate estimates of each error metric for each method; for censored data, the number is 960. 
\end{enumerate}

\subsection{Data Generation}
\label{sec4.1}

Our literature review revealed several important factors that contribute to the accuracy of estimates produced by various random forest methods. These factors include sample size, covariate type (continuous or categorical), number of covariates, strength of signal, and distribution of signal across covariates (whether all covariates are important or only a small fraction). For survival data, we added censoring rate as an additional factor. We refer to each simulation setting as a ``factor-level combination'' (FLC) since each setting is composed of a different combination of levels for these 5 (or 6) factors. We describe these factors and their levels below. 

In the uncensored setting, for each covariate value $\mathbf{X}=\mathbf{x}$, we simulated normally distributed responses with mean $\mathbf{x}^{\top}\boldsymbol{\beta}$ and standard deviation 1.2, where $\boldsymbol{\beta}$ is a vector of coefficients (excluding the intercept, which we take as 0). In the censored setting, we generated responses from a Weibull distribution with shape parameter $\rho=2.7$ and scale parameter $\lambda \exp\{\mathbf{x}^{\top}\boldsymbol{\beta}\}$, where $\lambda=0.8$. See Appendix~\ref{app:a} for more details on the covariates and coefficient values used in the simulations.

\paragraph{Sample Size.} For the uncensored data study, we selected three sample sizes for the training set: 300, 1200, and 2500. These values are approximately equally spaced on the square-root scale because the variability of many statistics is a function of $n^{-1/2}$. Due to the computational burden, we dropped the highest level for the censored-data study after observing little difference in the relative performance of different tuning methods between the medium and high levels in the uncensored-data study. The remaining two sample sizes reflect sizes of some survival time datasets we have encountered in practice. 

\paragraph{Covariate type.} For the uncensored-data study, we considered independent covariates that were all categorical, all continuous, or an equal mix of the two. For the censored data study, we only considered independent covariates that were either all categorical or all continuous. All categorical covariates had a multinomial distribution with either two or three categories, where the probability of each category ranged from 0.2 to 0.6, depending on the setting (see Appendix~\ref{app:a} for details). All continuous covariates had $U(0,1)$ distributions. 

\paragraph{Number of covariates.} For both studies, half of the FLCs comprised $p=4$ covariates, and the other half comprised $p=10$ covariates.

\paragraph{Signal-to-noise ratio.} A small-but-convincing literature demonstrates the relationship between the signal-to-noise ratio (SNR) and optimal RF tuning parameters (in particular, see \citealp{mentch2020}). Therefore, we selected three levels of the SNR, reflecting high, medium, and low levels. To quantify SNR for our normal responses, we used the standard SNR for linear regression \citep{friedman}, i.e.,
\begin{align}
    S\!N\!R = \frac{R^2}{1-R^2}.
    \label{snr}
\end{align}
For our Weibull responses, we first defined a pseudo $R^2$ as in \cite{berk2024},
\begin{align*}
    R_P^2 = \frac{\textrm{Var}(\textrm{E}(T|\mathbf{X}))}{\textrm{Var}(\textrm{E}(T|\mathbf{X})) + \textrm{E}(\textrm{Var}(T|\mathbf{X}))},
\end{align*}
which represents the proportion of the marginal variance of the survival times that results from differences in $X$. We then defined SNR as in \eqref{snr}, replacing $R^2$ with $R_P^2$. For each SNR value, we maintained this value across FLCs by simulating a large number of observations at each FLC, then adjusting $\boldsymbol{\beta}$ so that the SNR was approximately equal to the target value.

\paragraph{Distribution of signal.} Related to the SNR, we considered two levels of the distribution of signal across covariates: even and concentrated. For the even level, the coefficients associated with all covariates were non-zero. For the concentrated level, the coefficients were all zero except for one (when $p=4$) or two (when $p=10$), thus leaving the majority as noisy covariates. We included this factor based on findings from \cite{mentch2020} and \cite{berk2024} that suggested performance benefits in RFs and RSFs from the presence of noisy covariates; however, rather than add noisy covariates, we adjusted the distribution of signal as described---holding the SNR fixed---to change the number of important variables.

Refer to Appendix~\ref{app:a} for more specific details about the structure of the linear predictors---which depend on the covariate type, number of covariates, SNR, and distribution of signal.

\paragraph{Censoring rate.} For our censored-data study, we added the censoring rate as an additional factor, given the impact that censoring has on the performance of survival methods in general and previous findings that the censoring rate impacts the performance of survival forest methods in particular \citep{berk2024}. We consider two levels of censoring, 10\% and 30\%, representing light and moderate levels of censoring, respectively. We simulated censoring times to be independent with identical exponential distributions. The rate parameters were selected separately for each FLC to achieve the desired censoring proportion on average across datasets.

Tables~\ref{table_flcs_uncensored} and ~\ref{table_flcs_censored} summarize the factors and factor levels used in our simulation studies.

\begin{table}[!ht]
\begin{tabular}{p{0.3\textwidth} p{0.65\textwidth}}
\hline
\textbf{Factor}           & \textbf{Levels}                                 \\ \hline
Methods                   & QCL, MSPE, GRF, Default                         \\
$n$                       & 300, 1200, 2500                                  \\
Covariate Type            & Cat, Cont, CatCont                              \\
$p$                       & 4, 10                                            \\
SNR ($R^2$)               & High (0.75), Medium (0.50), Low (0.25)           \\
Signal Distribution       & Even, Concentrated                               \\ \hline
\end{tabular}
\captionsetup{justification=raggedright,singlelinecheck=false}
\caption{Factor levels used in our simulation study: uncensored data}
\label{table_flcs_uncensored}

\end{table}

\begin{table}[!ht]
\begin{tabular}{p{0.3\textwidth} p{0.65\textwidth}}
\hline
\textbf{Factor}           & \textbf{Levels}                                       \\ \hline
Methods                   & QCL-C, QCL-IPCW, C-index, Default                     \\
$n$                       & 300, 1200                                             \\
Covariate Type            & Cat, Cont                                             \\
$p$                       & 4, 10                                                 \\
SNR ($R^2$)               & High (0.75), Medium (0.50), Low (0.25)                \\
Signal Distribution       & Even, Concentrated                                    \\
Censoring                 & 10\%, 30\%                                            \\ \hline
\end{tabular}
\captionsetup{justification=raggedright,singlelinecheck=false}
\caption{Factor levels used in our simulation study: censored data}
\label{table_flcs_censored}
\end{table}

\subsection{Error metrics}
\label{sec4.2}

On each test dataset, we computed the estimated marginal bias and mean squared error (MSE) of both the quantile coverage probabilities and the quantile estimates. Among these four error metrics, the primary error metric, which our tuning procedure is tailored to minimize, is the coverage probability bias, but we included the others to check that gains in accuracy are not accompanied by large losses elsewhere.

For a given FLC and fitted RF, let $\tilde{\tau}_{i}=F_T (\hat{q}_{\tau}(\mathbf{x}_{i})|\mathbf{x}_{i})$ denote the true conditional coverage probability associated with the $\tau$-quantile estimate for observation $i=n+1,\ldots,N$. Then $\tilde{\tau}_{i} - \tau$ represents the conditional coverage bias of the quantile estimate at $\mathbf{x}_{i}$. We use the average of these quantities across the thousand observations in the test set to estimate the bias of the marginal coverage probability for the RF in question. Similarly, $\hat{q}_{\tau}(\mathbf{x}_{i})-q_{\tau}(\mathbf{x}_{i})$ represents the conditional bias of the quantile estimate at $\mathbf{x}_{i}$, and the average of these quantities across the test set is an estimate of the marginal bias of the quantile estimates for the RF in question.  

Low marginal bias may hide serious errors in conditional coverage probabilities that balance out fortuitously on average. We therefore also compute MSEs on both the coverage probability and quantile scales by squaring the conditional biases prior to averaging within the test set. A large MSE implies that the method has poor conditional properties, regardless of its marginal properties.   

To assess the accuracy of the different prediction interval methods we considered, we estimated coverage rates and widths evaluated on the 1080 test sets for all FLCs of each simulation study. To do so, we used the true conditional distribution of the quantile estimate for each test observation to compute the probability that an observation falls between the upper and lower quantile estimates, and then we averaged these probabilities across the test set.

\subsection{Implementation}
\label{sec4.3}
\noindent

All simulations were carried out using R, version 4.3.3. To fit RFs to our uncensored data, we used version 0.16.0 of the \texttt{ranger} package, which allows for a faster implementation of RFs than the \texttt{quantregforest} package, which relies on the \texttt{randomForest} package. For our ``default'' \texttt{mtry}, we used $p/3$ (the default in \texttt{quantregforest} and generally accepted default for regression analysis, instead of $\texttt{mtry}=\sqrt{p}$, the default in \texttt{ranger}). Our ``default'' \texttt{nodesize} was 5 (the default in both packages). With the \texttt{ranger} implementation, nodes may be split only if they are of \texttt{nodesize} or larger. To fit GRFs, we used version 2.3.2 of the \texttt{grf} package with its default tuning parameter values, $\texttt{mtry} = \min\left(\left\lceil \sqrt{p} + 20 \right\rceil, p\right)$ and $\texttt{nodesize}=5$. To fit RSFs to censored data, we used version 3.2.3 of the \texttt{randomForestSRC} package and used its default tuning parameters for our ``default'' setting: $\texttt{mtry}=\sqrt{p}$ and $\texttt{nodesize}=15$. In this package, \texttt{nodesize} represents the minimum number of failure times in a terminal node. Note that previous versions of the package used $\texttt{nodesize}=3$ as the default.

\section{Results}
\label{sec5}

We present our core results in the next two subsections. 

\subsection{Uncensored setting}
\label{sec5.1}

\subsubsection{Quantile estimation}
\label{sec5.1.1}

\begin{figure}[ht]
\centering
\includegraphics[width=1.0\textwidth]{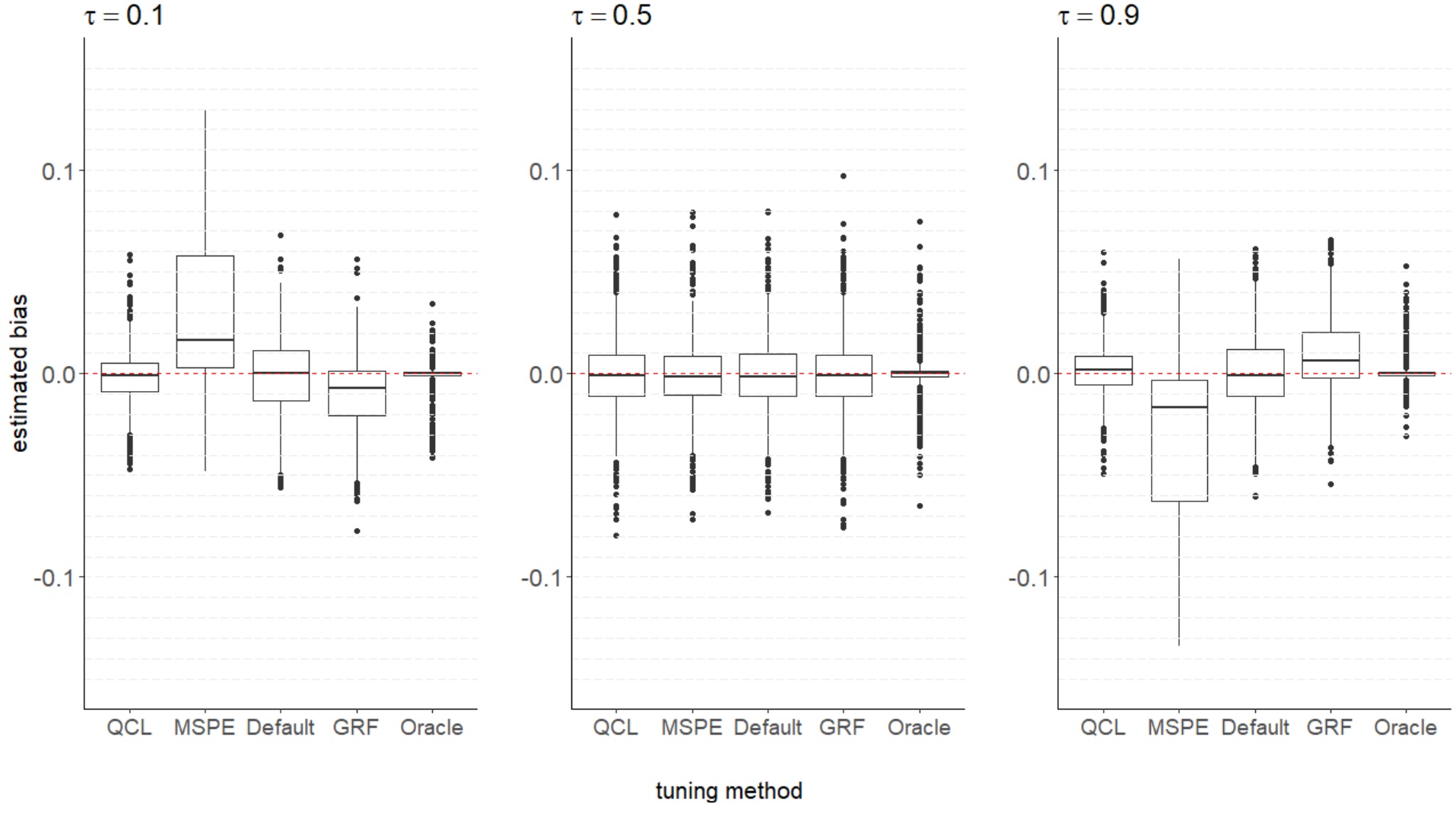}
\caption{Estimated marginal coverage probability bias vs.\ method for $\tau \in \{0.1, 0.5, 0.9\}$ in the uncensored data setting}
\label{fig_boxplots_covbias_uncensored}
\end{figure}

In Figure~\ref{fig_boxplots_covbias_uncensored}, we present boxplots of the estimated marginal coverage probability bias for the 0.1, 0.5, and 0.9 quantile estimates in the uncensored data setting. The plots summarize the distribution of the bias estimates from all 1080 datasets in the simulation study for each of the four methods we tested plus the Oracle (to show the best performance that could have been achieved from each dataset). Ideally, a method should produce quantile estimates with zero coverage bias; practically speaking, we seek methods whose marginal biases cluster tightly around 0. Simulation variability due to the finiteness of our test set was very low, resulting in standard errors of bias estimates that were generally within about 0.005. We note that the results for $\tau=0.9$ mirror those for $\tau=0.1$ due to the symmetry of the normal distribution; we thus discuss only the results for $\tau=0.1$ and $\tau=0.5$.

At $\tau=0.1$, the distribution of estimated marginal coverage probability bias for QCL tuning is centred at 0 and has the least variability among RF methods (except the Oracle). Using default tuning parameters also results in estimated marginal coverage biases that are centred at 0, but these biases are more variable than those produced by QCL tuning. Traditional MSPE tuning produces 0.1-quantile estimates whose marginal coverage is generally too high, and the overall performance is erratic compared to that of other methods. The GRF-based 0.1-quantile estimates tend to under-cover. 

In contrast, for the 0.5 quantile, all methods produce very similar distributions of estimated marginal coverage probability bias. Interestingly, for all the values of $\tau$, there were many datasets for which even the Oracle RF produced quantile estimates with somewhat biased coverage. We address this point below. 

\begin{figure}[ht]
\centering
\includegraphics[width=1.0\textwidth]{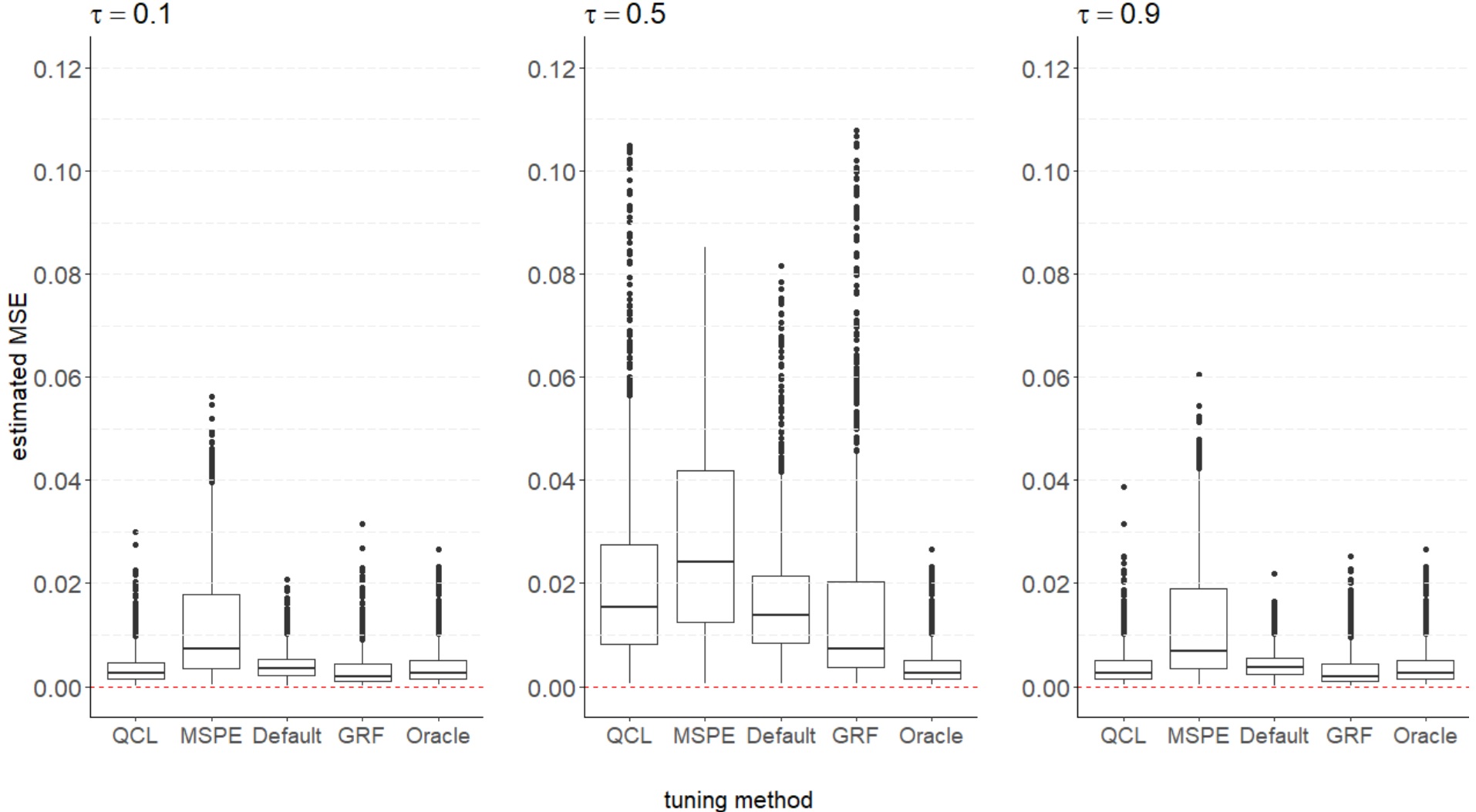}
\caption{MSE of estimated marginal coverage probabilities vs.\ method for $\tau \in \{0.1, 0.5, 0.9\}$ in the uncensored data setting}
\label{fig_boxplots_covmse_uncensored}
\end{figure}

In Figure~\ref{fig_boxplots_covmse_uncensored}, we present boxplots of the MSE of the estimated marginal coverage probabilities across datasets. For $\tau=0.1$ and $\tau=0.9$, all methods except MSPE tuning produced distributions of MSE that were similar to those produced by the Oracle. MSPE tuning typically produced larger MSEs than did the other methods. On the other hand, for $\tau=0.5$, QCL tuning and the default method led to similar values of the median MSE, while GRF led to a lower median MSE. MSPE tuning led to relatively high median MSE, which is surprising given that the true quantiles coincide with the means of the response in this case (and the estimated mean response is the optimal predictor of the response when MSE is used as the loss function). We explain this finding and highlight the importance of tuning to a specific target in the Discussion. 

\begin{figure}[ht]
\centering
\includegraphics[width=1.0\textwidth]{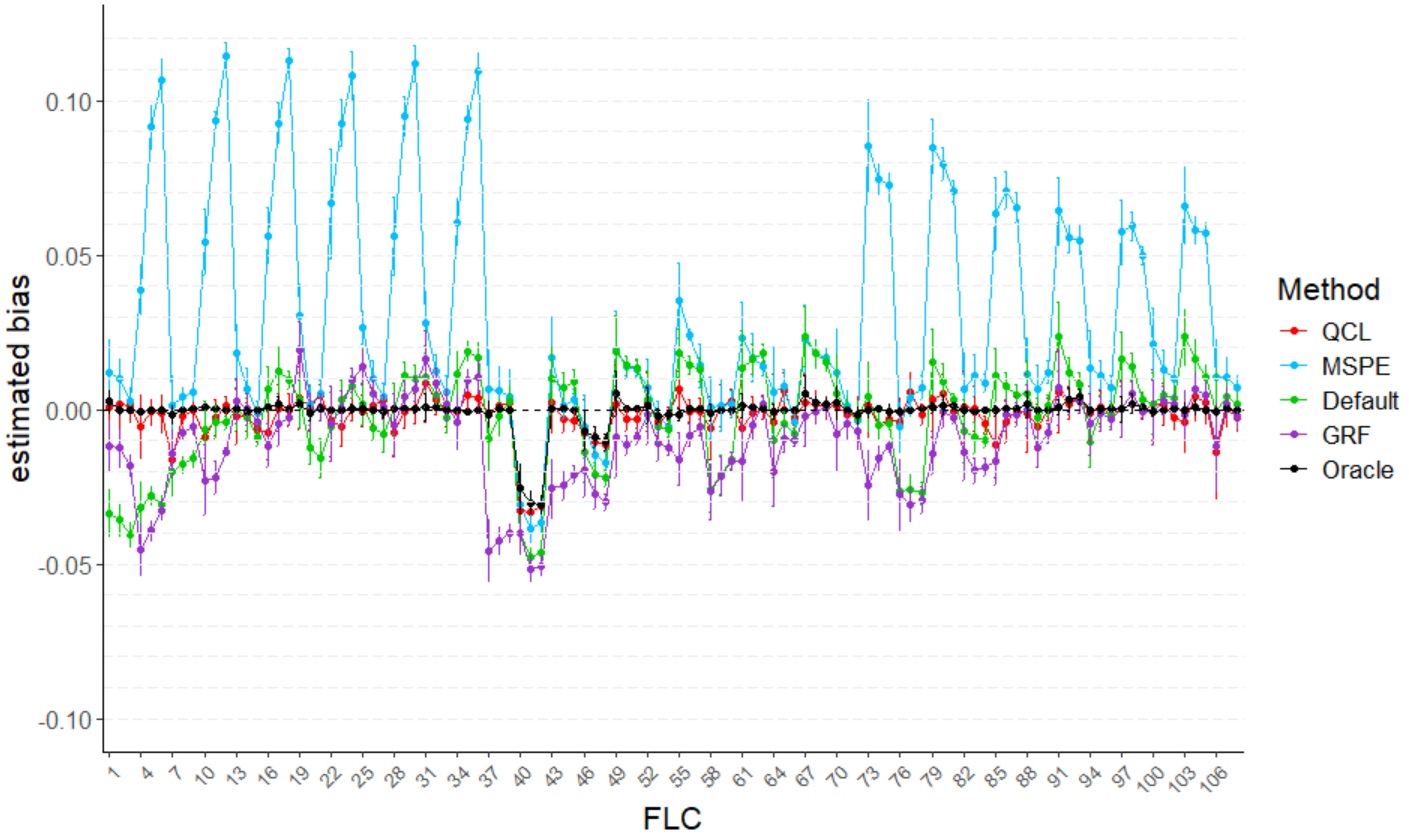}
\caption{Estimated coverage probability bias vs.\ FLC for each method ($\tau=0.1)$}
\label{fig_lineplots_covbias_uncensored}
\end{figure}

To explore our results further, Figure~\ref{fig_lineplots_covbias_uncensored} shows the estimated marginal coverage probability bias (with confidence intervals) vs.\ FLC for the 0.1 quantile. The plotted points for each method are the sample means of the estimated marginal coverage probability biases based on the 10 datasets for each FLC. The actual factor levels corresponding to each FLC are given in Appendix~\ref{app:b}. We focus on the 0.1 quantile because the plot for the 0.9 quantile mirrors Figure~\ref{fig_lineplots_covbias_uncensored}, and all methods performed similarly when estimating the 0.5 quantiles.

An obvious feature of Figure~\ref{fig_lineplots_covbias_uncensored} is that most methods produced quantile estimates with coverage probabilities that exhibit significant marginal bias for most FLCs. Only QCL tuning and the Oracle succeeded in providing reliable, near-zero mean marginal coverage bias. Specifically, out of 108 FLCs, the confidence intervals for mean marginal bias covered zero in 97 settings in the case of the Oracle, 95 settings for QCL tuning, 43 for GRF, 37 for the Default, and 30 for MSPE tuning.

MSPE tuning, in particular, led to serious over-coverage in its 0.1-quantile estimates in many settings, providing one example of the consequences when the tuning target is very different from the goal. (In this case, the tuning target is an estimated mean with low MSPE, while the goal is estimating a tail quantile with accurate coverage probability.) We also see that GRF often led to substantial under-coverage in its quantile estimates.   

In several settings (most prominently, FLCs 40--42), even the Oracle RF is incapable of estimating 0.1 quantiles without substantial marginal coverage bias. These FLCs correspond to data structures with the highest complexity: 10 continuous covariates, high SNR, and an even distribution of signal (for the three sample sizes tested). These three settings combine levels of factors that make the response surface more difficult to approximate. In general, infinitely many terminal nodes would be required for a forest to accurately represent an FLC when at least one covariate is continuous. When a strong signal is spread evenly across many continuous covariates, finite terminal nodes can approximate the surface only very crudely. Thus, even the Oracle RF is unable to achieve good estimates, reflecting a general shortcoming of RFs rather than of any tuning procedure.

We ran some additional simulations to explore this problem further, separately increasing each of SNR, $p$, and $n$, while holding all other factor levels constant. Increasing SNR or increasing $p$ while holding other factor levels constant led to more biased estimated coverage probabilities. Increasing $n$ to allow larger trees to be built slightly reduced the bias, but a substantial increase in $n$ was required to achieve even a slight improvement (e.g., $n=50,000$ resulted in only about a 0.003 reduction in coverage probability bias compared to our $n=2,500$ setting).

We also conducted a limited supplementary study to investigate whether estimates of the population QCL were inaccurate (thus causing coverage bias in the test set). We found that QCL estimates were very accurate. (See Appendix~\ref{app:f} for details.)

In Appendix~\ref{app:c}, we present a plot that is similar to that in Figure~\ref{fig_lineplots_covbias_uncensored} but that shows estimated MSE vs.\ FLC for each method. In Appendix~\ref{app:e}, we provide the analogous plot to Figure~\ref{fig_lineplots_covbias_uncensored} showing the estimated marginal bias in the quantile estimates (rather than their associated coverage probabilities). Although QCL tuning is designed to minimize bias in coverage probability, these plots provide reassurance that the method also performs competitively with respect to other metrics that are relevant to quantiles.

\subsubsection{Prediction intervals}
\label{sec5.1.2}

\begin{figure}[ht]
\centering
\includegraphics[width=1.0\textwidth]{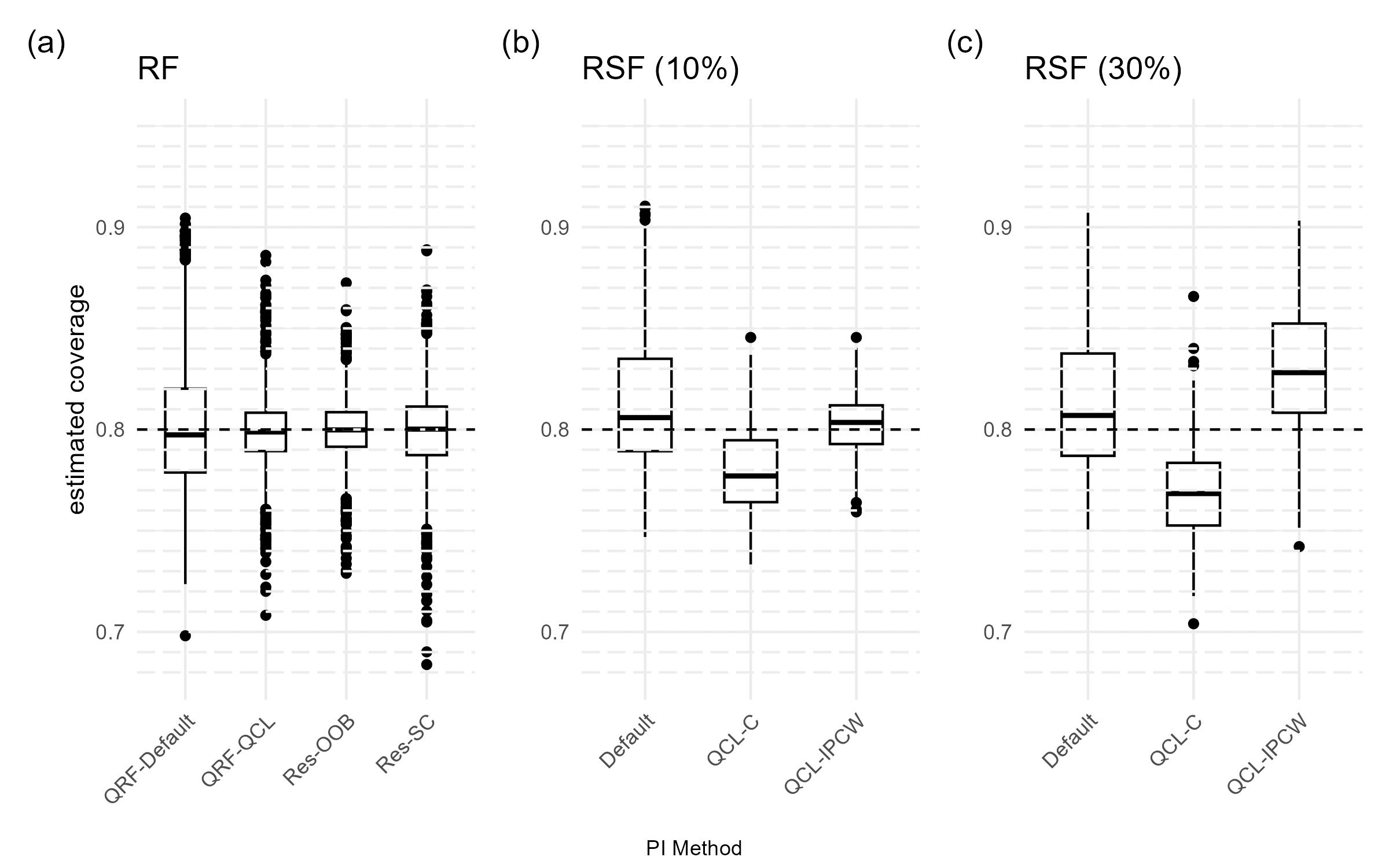}
\caption{Marginal coverage rates for prediction intervals in the (a) uncensored setting, (b) 10\% censoring setting, and (c) 30\% censoring setting. Plots (b) and (c) are based on the data from the $n=1200$ setting only.}
\label{fig_boxplots_picoverage}
\end{figure}

Figure~\ref{fig_boxplots_picoverage}(a) displays boxplots of the marginal coverage rates for prediction intervals in the uncensored-data setting. These plots summarize the distribution of coverage rates for all 1080 datasets for each of the four methods for constructing prediction intervals described in Section~\ref{sec4}: QRF-Default, QRF-QCL, Res-OOB, and Res-SC. See Appendix~\ref{app:d} for descriptive statistics associated with the coverage rates and interval widths and for plots of coverage rates broken down by each FLC. All intervals have similar distributions of coverage rates, except QRF - Default intervals, which have greater variability. The same can be said for their interval widths. These results indicate that QCL tuning is successful in improving the properties of base QRF intervals.

Previous investigations \citep{zhang, roy} found that QRF intervals generally performed worse than alternative methods in terms of both coverage rate and width, but their studies were based on either untuned \citep{roy} or suboptimally tuned \citep{zhang} forests. Our results demonstrate that tuning QRF intervals to the QCL target nullifies previously reported advantages for both Res-OOB and Res-SC intervals. 

Res-OOB and Res-SC intervals are based on the assumption that the error distribution is the same for all $\mathbf{x}$ (a limitation of these methods). Furthermore, the versions of Res-OOB and Res-SC that we implemented assumed symmetry of the intervals. Both of these assumptions hold in our simulation studies, where we have generated normally distributed errors. But we need to be mindful that we have effectively created very favourable settings under which to assess their relative performance. Nonetheless, QCL-tuned intervals perform comparably to Res-OOB and Res-SC intervals in terms of both coverage and width.

In contrast, QRF intervals do not make assumptions about the distribution of the errors. We therefore might expect intervals based on targeted QCL tuning to outperform the other intervals when the assumptions underlying the latter are violated. The study by \cite{zhang} (though based on MSPE-tuned, not QCL-tuned, QRF intervals) provide some support for this hypothesis: while MSPE-tuned intervals were more likely than Res-OOB and Res-SC intervals to exhibit over- or under-coverage marginally and to have much more variable widths, they were more likely to produce improved conditional coverage in heterodscedastic settings. 

We did investigate whether tuning RFs separately for the two interval endpoints was necessary. We found that intervals produced by a singly tuned RF performed substantially worse than those produced via the approach that we developed in Section~\ref{sec3.2} and used in the simulation studies.

\subsection{Censored setting}
\label{sec5.2}

\subsubsection{Quantile estimation}
\label{sec5.2.1}

\begin{figure}[ht]
\centering
\includegraphics[width=1.0\textwidth]{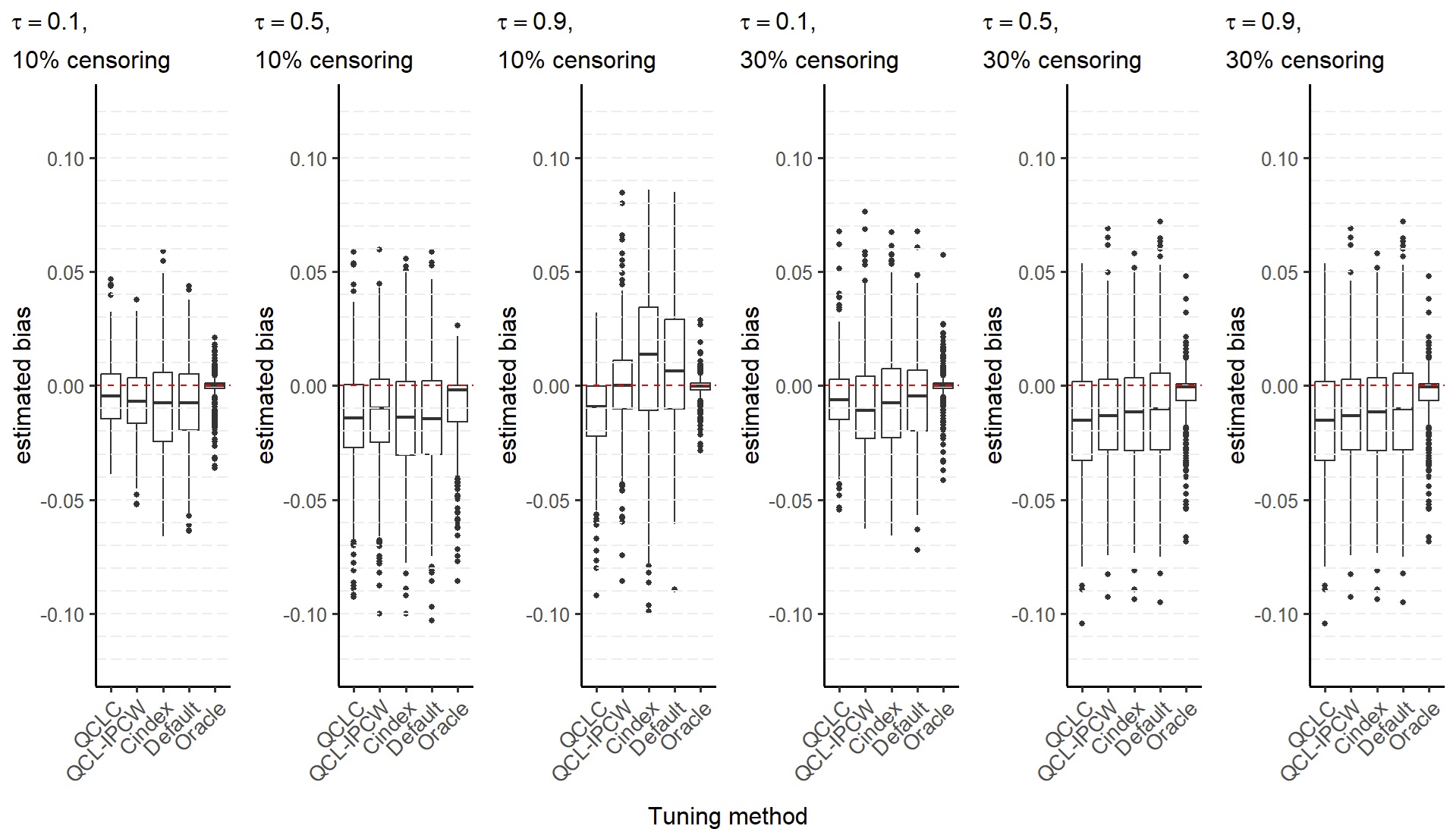}
\caption{Estimated coverage probability bias vs.\ method for $\tau \in \{0.1, 0.5, 0.9\}$ --- censored-data setting, $n=1200$}
\label{figboxplots_covbias_censored}
\end{figure}

Figure~\ref{figboxplots_covbias_censored} presents boxplots of the estimated coverage probability bias in the censored data setting, broken down by censoring level. The plots illustrate the distribution of bias estimates for all 960 datasets for each method we tested, plus the Oracle. Here we comment on the findings for all three quantiles because the data were simulated from (asymmetric) Weibull distributions. The greater variability in average biases across FLCs makes it difficult to detect performance differences across methods.

\begin{figure}[ht]
\centering
\includegraphics[width=1.0\textwidth]{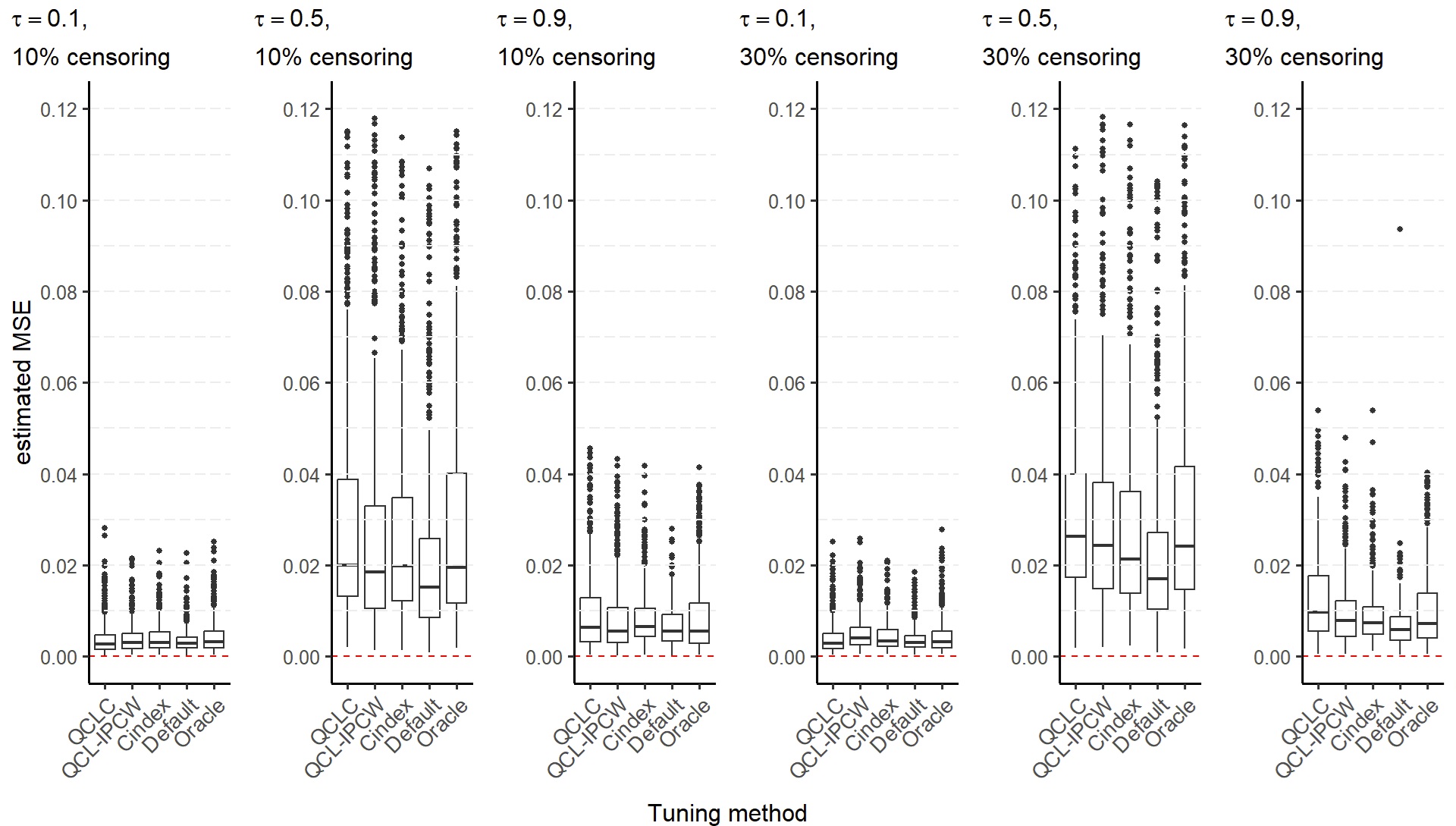}
\caption{Estimated MSE of coverage probabilities vs.\ method for $\tau\in \{0.1, 0.5, 0.9\}$ --- censored-data setting, $n=1200$}
\label{figbxoplots_covmes_censored}
\end{figure}

Figure~\ref{figbxoplots_covmes_censored} displays the estimated MSEs of coverage probabilities in the censored setting. Similar to the uncensored setting, QCL tuning led to estimated MSE values that were, on average, comparable to those produced using default tuning parameters and tuning via C-index. Using default parameters seems to offer a slight reduction in variability of conditional coverage, even better than the Oracle, which chooses the parameters that minimize marginal bias in each test set, not the MSE.  

\begin{figure}[!htbp]
\centering
\includegraphics[width=1.0\textwidth]{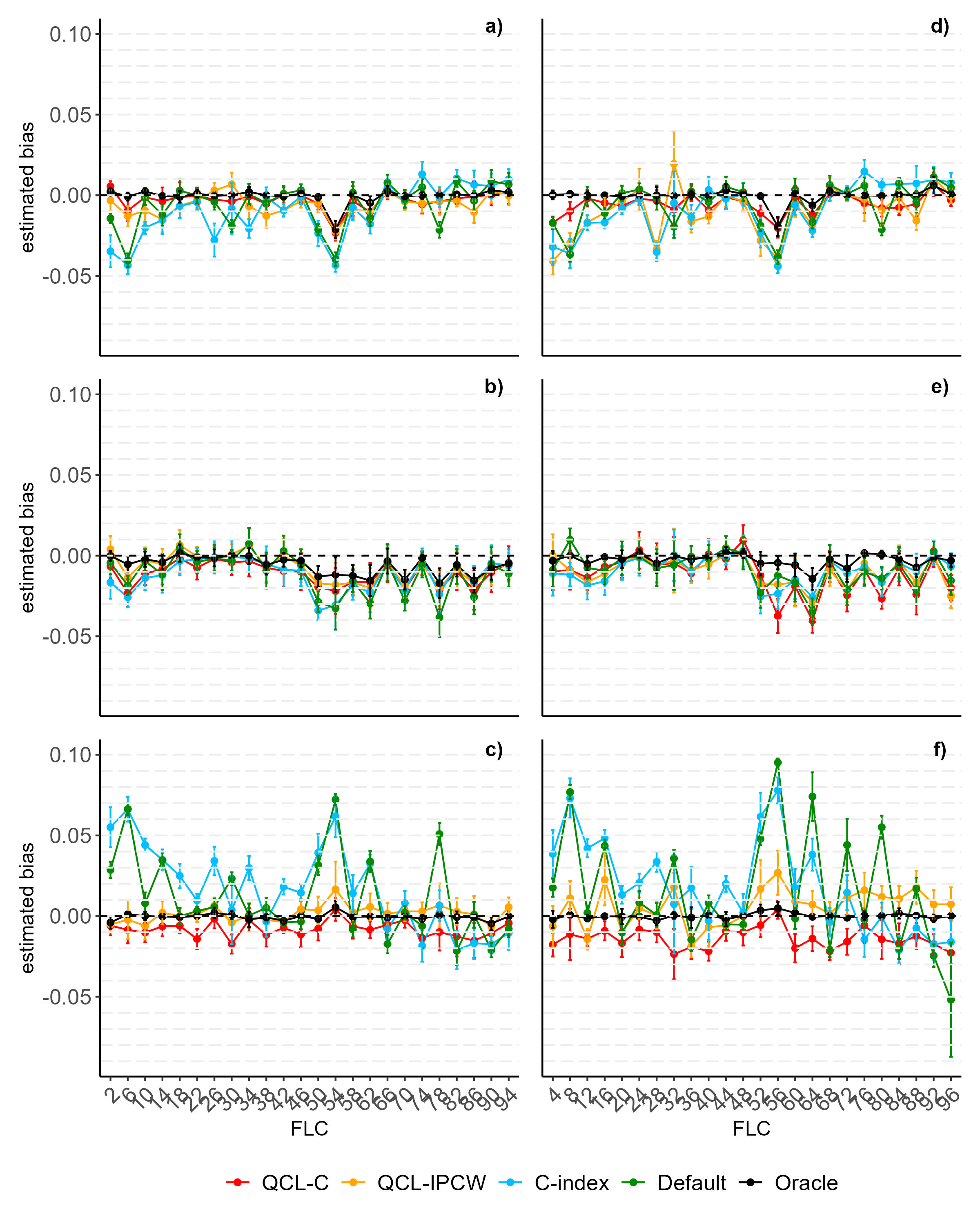}
\caption{Estimated coverage probability bias vs.\ FLC for each method at $n=1200$. Plots a), b), and c) correspond to the $\tau \in \{0.1, 0.5, 0.9\}$ quantiles, respectively, in the 10\% setting; plots d), e), and f) correspond to the $\tau \in \{0.1, 0.5, 0.9\}$ quantiles, respectively, in the 30\% setting. The estimated biases exclude observations where the quantile estimate did not exist.}
\label{figrsflineplotscombined}
\end{figure}

Figure~\ref{figrsflineplotscombined} displays plots of the estimated coverage probability biases, broken down by each FLC and separated by the censoring rate for the higher sample-size setting. As we did for the uncensored setting, the sample means from the 10 datasets for each FLC are plotted for each method along with the Oracle, with bars representing 95\% confidence intervals. These plots demonstrate that the estimated coverage probabilities associated with using default tuning parameters and C-index tuning tend to be systematically low for $\tau=0.1$ and systematically high for $\tau=0.9$ when using more conventional approaches. With a few exceptions---mostly for the 30\% censoring scenario---QCL-C and QCL-IPCW tend to produce estimated quantiles with less coverage probability bias. Moreover, even for FLCs where the CIs for QCL-C and QCL-IPCW did not cover 0, the bias estimates were typically closest to those produced by the Oracle, thus demonstrating that these methods are an improvement over Default and C-index. We concentrate on the $n=1200$ setting because the estimates are more precise and therefore the trends are more apparent, but the results are similar in the $n=300$ setting. Also note that in the 30\% censoring setting, the 0.5- and especially the 0.9-quantile estimates were frequently undefined because the estimated survival function did not drop down far enough (see Section~\ref{sec3.3}), in which case the observations were omitted when estimating the errors.

\subsubsection{Prediction intervals}
\label{sec5.2.2}

In the censored setting, one-sided intervals may be preferred. We don't provide separate plots for one-sided intervals because the coverage rates of the estimated 0.1 and 0.9 quantiles could also be thought of individually as lower and upper bounds, respectively, for 90\% prediction intervals. The results cumulatively suggest benefits of QCL tuning for both one- and two-sided intervals. 

However, we also show and briefly discuss results for two-sided intervals. Figures~\ref{fig_boxplots_picoverage}(b) and (c) summarize prediction interval coverage rates for each of the 960 censored datasets in the $n=1200$ setting, separately for the two censoring levels. At both censoring levels, QCL-IPCW and Default produced valid intervals, on average, whereas QCL-C did not (because the upper quantile estimates produced by QCL-C tended to be biased slightly low, whereas those produced by QCL-IPCW and Default tended to be biased high). QCL-IPCW intervals had the lowest variability in coverage rates across different settings, followed by QCL-C and then Default intervals. (Again, see Appendix~\ref{app:d} for more details.) 

QCL-IPCW and Default intervals had comparable mean widths in the 10\% censoring setting. In the 30\% censoring setting, QCL-IPCW intervals had similar mean widths as Default intervals but drastically larger median widths due differences in skewness. In both censoring settings, QCL-IPCW intervals had lower standard deviations in widths than Default intervals (see Table~\ref{table_picovwidth_uncensored} in Appendix~\ref{app:d}).

\subsection{Additional results: optimal tuning parameter values}
\label{sec5.3}

\begin{figure}[ht]
\centering
\includegraphics[width=1.0\textwidth]{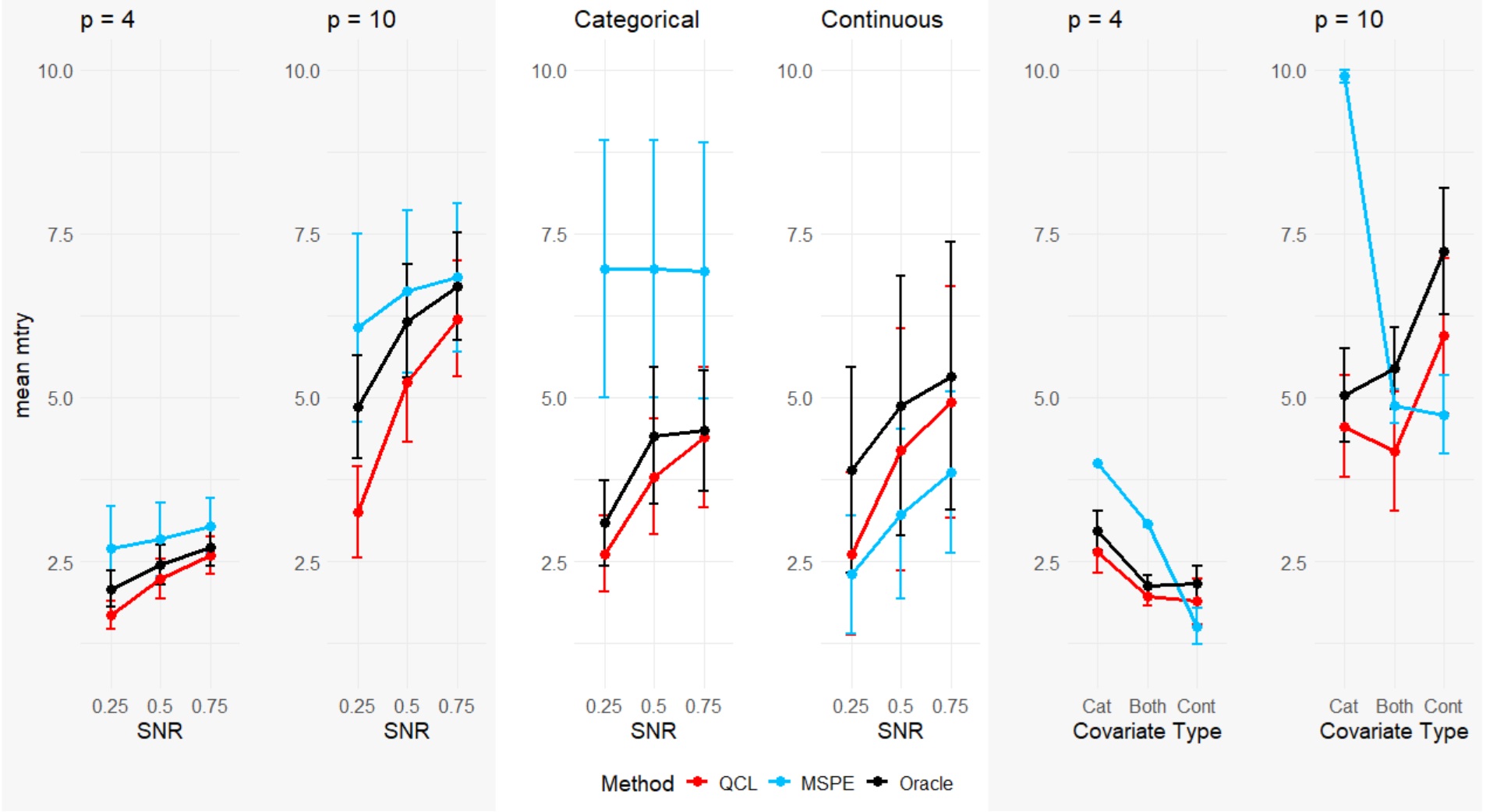}
\caption{Mean optimal \texttt{mtry} for QCL tuning, MSPE tuning, and the Oracle when estimating the $0.1$ quantile.
Left pair of plots: Mean \texttt{mtry} vs. SNR for $p=4$ and $p=10$.
Centre pair of plots: Mean \texttt{mtry} vs.\ SNR in the all‐categorical and all‐continuous covariate settings.
Right pair of plots: Mean \texttt{mtry} vs.\ covariate type (all categorical, both categorical and continuous, all continuous) for $p=4$ and $p=10$.}
\label{figmtry}
\end{figure}

The importance of tuning is amplified by the substantial variation we observed in optimal tuning parameter combinations across FLCs. Figure~\ref{figmtry} displays the average optimal \texttt{mtry} determined by QCL-tuning, MSPE-tuning, and the Oracle when estimating the $0.1$ quantile. The plots show the effect of SNR, covariate type, and $p$ on this value. 

Several trends are apparent. First, the higher the SNR, the higher the value of \texttt{mtry} determined by the Oracle. This result replicates a major finding in \cite{mentch2020}, which found a robust, positive relationship between SNR and \texttt{mtry} in the standard random forest setting (i.e., when point prediction was the goal and MSE was the evaluation metric). Our work empirically confirms that this relationship holds when estimating tail quantiles and when the goal is minimizing the absolute coverage bias. Tuning with QCL also results in a positive relationship between SNR and \texttt{mtry}. Importantly, the value of \texttt{mtry} obtained via this method is often slightly less than optimal. Tuning by MSPE leads to \texttt{mtry} values that are further away from those determined by the Oracle than tuning by QCL, especially when all covariates are categorical.  

Second, the covariate type affects the optimal \texttt{mtry}, but the direction of this relationship depends on the number and type of covariates. For $p=4$, the optimal \texttt{mtry} was \textit{lower} with all continuous covariates compared to categorical. For $p=10$, the optimal \texttt{mtry} was \textit{higher} with all continuous covariates compared to categorical. Tuning with QCL again mimicked this pattern much more closely than did traditional MSPE tuning. These results suggest that comprehensive study of the relationship between optimal tuning parameters and data properties would be worthwhile and could improve tuning efficiency by suggesting narrower ranges for optimal tuning parameter values.

\begin{figure}[ht]
\centering
\includegraphics[width=1.0\textwidth]{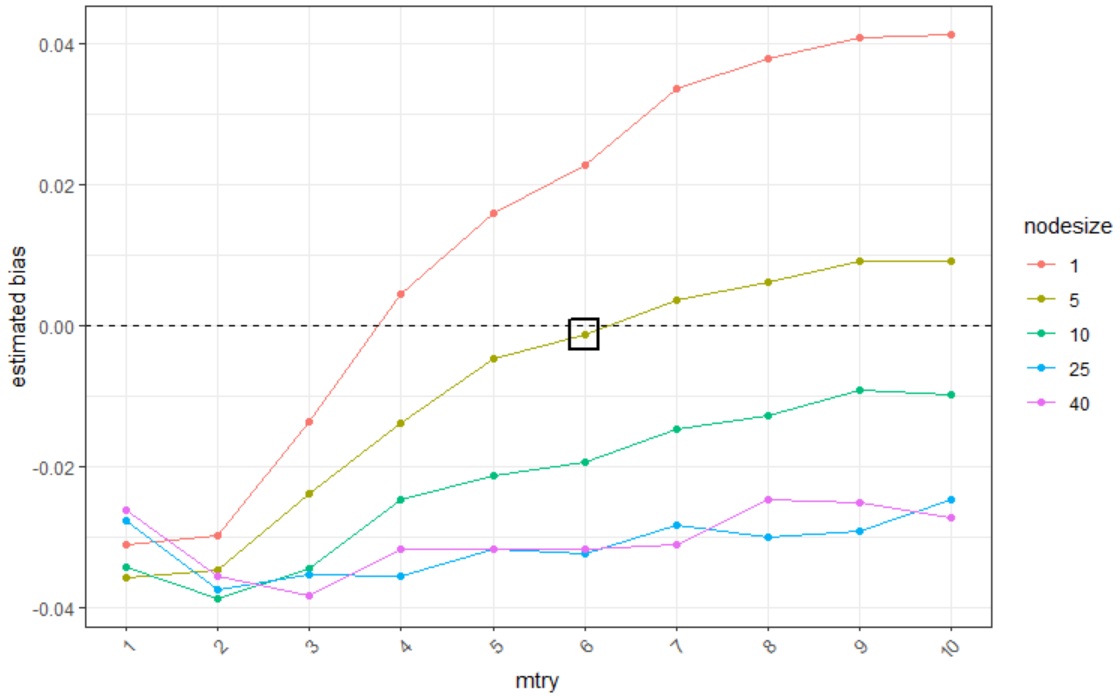}
\caption{Estimated coverage probability bias vs.\ \texttt{mtry} for various values of \texttt{nodesize} and one select FLC ($\tau=0.1)$; the squared point denotes the optimal combination selected by QCL, which coincides with the oracle in this setting}
\label{figtuningvsbias}
\end{figure}

Figure~\ref{figtuningvsbias} demonstrates that \texttt{mtry} and \texttt{nodesize} are similarly important for minimizing quantile coverage probability bias. Furthermore, there is some effect of interaction between the two tuning parameters. For example, when \texttt{nodesize} is small, changing \texttt{mtry} has a relatively large effect on the bias; conversely, when \texttt{nodesize} is large, \texttt{mtry} has a much smaller effect. Likewise, at large values of \texttt{mtry}, different \texttt{nodesize} values have a relatively large effect on the bias, whereas at small \texttt{mtry}, the impact of \texttt{nodesize} is minimal. Because these two tuning parameters interact, tuning one in isolation (e.g., tuning only \texttt{mtry}) can miss regions where their joint effect achieves the least bias. Thus, in practical terms, the implication is that tuning both \texttt{mtry} and \texttt{nodesize} in tandem is valuable.

\section{Discussion}
\label{sec6}

In the uncensored data setting, tuning via QCL led to coverage probability estimates for tail quantiles with no evidence of bias in the vast majority of FLCs we tested. All alternative methods often produced highly biased estimates. Therefore, our recommendation in this setting is straightforward: use QCL tuning when estimating more extreme quantiles and for constructing prediction intervals. The costs of not tuning or conventional tuning can be high, depending on the underlying data structure. On the other hand, there is no need to tune the median. 

In the setting where data were lightly censored, tuning via QCL-C and QCL-IPCW led to much less biased coverage probability estimates of tail quantiles compared to alternative methods. With moderate censoring, tuning with QCL-C led to substantially less biased estimates of lower quantiles than tuning conventionally or using the default tuning parameter values, and both QCL-C and QCL-IPCW most reliably led to the least biased estimates of upper quantiles. With light censoring, we recommend tuning with either QCL-C or QCL-IPCW for all quantiles, especially upper quantiles. With more moderate censoring, tuning QCL-C seems beneficial when estimating lower quantiles, while tuning with either QCL-C or QCL-IPCW is recommended when estimating upper quantiles. However, tuning when estimating middle quantiles may lead to more modest or negligible benefits.  For upper one-sided intervals, we recommend tuning with either QCL-C or QCL-IPCW, depending on the censoring rate. For two-sided prediction intervals, we recommend tuning using QCL-IPCW, which, unlike QCL-C, tends to produce valid intervals and, compared to Default, tends to produce less variable interval coverage and widths.

This work gives one example of the importance of aligning the tuning method with the estimation goals. Even small changes in loss functions can impact the performance of parameter estimates. For example, even though QCL tuning targets the absolute coverage bias of quantile estimates and achieved more accurate estimates in terms of the most closely related evaluation metric, QCL tuning did not achieve clearly superior estimates in terms of other metrics. Indeed, even the Oracle, which was clearly superior to all empirical tuning approaches with respect to the metric based on the targeted loss, performed no better than empirical tuning according to other error metrics. Therefore, if we want to achieve optimality according to one particular metric, even subtle changes in the loss function---such as using squared errors (MSPE) rather than absolute bias---can have a substantial impact.

On a related note, as we discussed in Section~\ref{sec5.1}, it may seem surprising that MSPE tuning produced median estimates with higher estimated MSE than did the default, which seems to contradict conventional wisdom that MSPE tuning improves RF predictions \citep{probst}. However, the standard goal for RFs is to produce accurate point predictions using estimated means, not medians. While these parameters are the same in normal distributions (including those that we used to generate the data), their RF estimates are different. When we instead tuned using a version of OOB MSPE that uses medians instead of means as predicted values, the estimated MSE decreased---and was lower than that attained using default tuning parameters, QCL tuning, or GRFs. This finding furthers bolsters our larger point that RFs should be tuned according to the estimation target. 

We used marginal, rather than conditional, coverage probability bias as our tuning target because marginal evaluation coverage probability bias is both important and feasible. Guaranteeing accurate conditional coverage probability at each covariate value would be ideal, but it is not clear how to use tuning to allow for accurate estimation of conditional coverage for each value in the covariate space simultaneously. Focusing on marginal coverage is a reasonable compromise. However, selecting forests to achieve good marginal coverage does not guarantee low variability in the conditional coverage. If the goal is to minimize the conditional bias across the full range of the covariate space, an entirely different approach may be required.

This paper focuses on the finite-sample performance of RF-based methods, where tuning can have an important impact on the accuracy of parameter estimates. In the asymptotic setting, random forests have been shown to produce consistent estimates of the mean response under certain (typically restrictive) assumptions about the RF algorithm, data structure, and tuning parameters. Importantly, existing theoretical results \citep{breiman2004, ishwaran2010, biau2010, biau2012, denil, wager, scornet2015, biauscornet, meinshausen, edcosaque} do not directly address the practical implications of tuning for finite-sample performance. 

Few other papers in the literature investigate the estimation of tail or extreme quantiles using random forests. Only one paper to date has focused on estimating extreme quantiles via random forests \citep{gnecco}. Their approach looked at extreme quantiles (targeting values of $\tau$ very close to 1) beyond the observed response range given a set of covariates using tail approximations motivated by extreme value theory and tailored primarily to heavy-tailed distributions. Their method, called extremal random forests, involves estimation of the parameters of a generalized Pareto distribution by maximizing a local likelihood, with weights extracted from a RF.

In the censored data setting, we used RSF because it is the most popular and well-studied among survival forest methods, its implementation is computationally efficient, and preliminary results suggested that targeted tuning is more important than the specific forest implementation used. However, tuned versions of other survival forests could perform better than RSFs. In fact, our earlier investigation found that some other variants may be preferable when the goal is estimating a survival function or making a point prediction \citep{berk2024}, but this comparison used default tuning parameters. Moreover, code is not publicly available for many of these alternative survival forests, and the code that we obtained does not easily allow for extraction of important quantities (e.g., OOB information).

Some limitations of the methods we used include the difficulty in estimating quantiles with censored data (specifically, estimating middle or upper quantiles with heavier censoring), the potentially high computational time to tune with large datasets and/or many covariates, the persistence of conditional bias despite optimizing for marginal bias, and the inherent limitations of RF estimates due to the way the ECDF is constructed. Moreover, all major RF packages allow ``pure" variables---variables that cannot be used for splitting because all values are the same in a node---to be selected into the pool as splitting candidates. If such a variable is selected for splitting, the splitting terminates (prematurely), ultimately inflating bias in RF estimates.

Limitations of our study include the finite number of settings in which we tested our tuning method, our restricted focus on three quantiles, our choice to consider only a finite, fixed set of possible tuning parameters values when tuning, and our use of grid search to tune (different approaches might prove to be more effective, \citealp{bayley}). Though our paper relates to targeted tuning in general, we focused on quantile estimation and the development of a loss function for this purpose. To tune RFs to achieve other estimation goals, different loss functions may need to be developed. Moreover, we limited our attention to targeted tuning in RFs and RSFs. We have not investigated how targeted tuning may interact with other forest methods. In particular, GRFs use a splitting criterion that is more closely tailored to a specific estimation problem than squared-error loss, analogous to the use of specialized loss functions for tuning. It would be interesting to compare these techniques across a wider range of problems and to see whether targeted tuning can improve GRFs. These limitations---in both the methods we used and our study---all present opportunities for future research.

In summary, we have proposed a novel tuning procedure that demonstrates the importance of tuning using a loss function that closely aligns with one's estimation or prediction goal. For the goal of estimating quantiles, we compared our tuning procedure using fully observed data and censored data to other methods and found clear benefits of targeted tuning for quantile estimation and, relatedly, prediction intervals. Our work paves the way for the development of other tuning procedures that match a variety of goals in the context of both random forests and other machine learning methods.


\acks{This work was supported, in part, by two Natural Sciences and Engineering Research Council of Canada grants (RGPIN-04304-2018 and RGPIN-2024-05146).}


\newpage

\appendix
\section{Linear predictors used}
\label{app:a}

This appendix details the coefficient vectors and categorical covariate distributions used in the simulation study. These specifications fully determine the linear predictors, $\mathbf{X}^{\top}\boldsymbol{\beta}$, used to generate responses. For each covariate type, we consider two settings for the number of covariates ($p=4$ and $p=10$) and two configurations for the distribution of signal across covariates (``even'' vs.\ ``concentrated''). In addition, the coefficients in the high SNR setting are provided, and the coefficients in the medium and low SNR settings are defined as scalar multiples ($\boldsymbol{\beta}_{\text{med}}$ and $\boldsymbol{\beta}_{\text{low}}$, respectively) of the high SNR coefficients.

\subsection{Coefficient vectors}
\label{app:a.1}

Tables~\ref{tab:beta_cat}--\ref{tab:beta_catcont} provide the high SNR coefficient vectors for the three covariate types.

\begin{table}[!htbp]
\centering
\begin{tabular}{lp{0.75\textwidth}}
\toprule
Setting & High SNR $\boldsymbol{\beta}$ vector \\ \midrule
Even signal, $p=4$   & $\boldsymbol{\beta}_{1} = 0.64 \times (5,\; 4,\; -2.5,\; -2.3)$ \\[1ex]
Even signal, $p=10$  & $\boldsymbol{\beta}_{2} = 0.445 \times (5,\; 4,\; -2.5,\; -2.3,\; -3,\; 2,\; -1.5,\; 1,\; -2,\; -3,\; 2.2,\; -0.8,\; 2.5)$ \\[1ex]
Concentrated signal, $p=4$  & $\boldsymbol{\beta}_{3} = (4.5,\; 0,\; 0,\; 0)$ \\[1ex]
Concentrated signal, $p=10$ & $\boldsymbol{\beta}_{4} = (4.4,\; 1,\; 0,\; 0,\; 0,\; 0,\; 0,\; 0,\; 0,\; 0,\; 0,\; 0,\; 0)$ \\ \bottomrule
\end{tabular}
\caption{High SNR coefficient vectors for \textbf{categorical} predictors. The scalar multiples in the medium and low SNR settings are  
$\boldsymbol{\beta}_{1,\text{med}} = 0.575\,\boldsymbol{\beta}_{1}$, $\boldsymbol{\beta}_{2,\text{low}} = 0.33\,\boldsymbol{\beta}_{2}$,
$\boldsymbol{\beta}_{3,\text{med}} = 0.58\,\boldsymbol{\beta}_{3}$, and $\boldsymbol{\beta}_{4, \text{low}} = 0.335\,\boldsymbol{\beta}_{4}$.}
\label{tab:beta_cat}
\end{table}

\begin{table}[!htbp]
\centering
\begin{tabular}{lp{0.75\textwidth}}
\toprule
Setting & High SNR $\boldsymbol{\beta}$ vector \\ \midrule
Even signal, $p=4$   & $\boldsymbol{\beta}_{5} = (5,\; 4,\; -2.5,\; -3.7)$ \\[1ex]
Even signal, $p=10$  & $\boldsymbol{\beta}_{6} = 0.73 \times (5,\; 4,\; -2.5,\; -4,\; -5,\; 0.5,\; 1.5,\; -3,\; 3,\; 2.5)$ \\[1ex]
Concentrated signal, $p=4$  & $\boldsymbol{\beta}_{7} = (7.8,\; 0,\; 0,\; 0)$ \\[1ex]
Concentrated signal, $p=10$ & $\boldsymbol{\beta}_{8} = (7.6,\; 2,\; 0,\; 0,\; 0,\; 0,\; 0,\; 0,\; 0,\; 0)$ \\ \bottomrule
\end{tabular}
\caption{High SNR coefficient vectors for \textbf{continuous} predictors. The medium and low SNR coefficients are defined as $\beta_{5,\text{med}} = 0.58\,\boldsymbol{\beta}_{5}$, $\boldsymbol{\beta}_{6,\text{low}} = 0.335\,\boldsymbol{\beta}_{6}$,
$\beta_{7,\text{med}} = 0.575\,\boldsymbol{\beta}_{7}$, and
$\boldsymbol{\beta}_{8,\text{low}} = 0.335\,\boldsymbol{\beta}_{8}$.}
\label{tab:beta_cont}
\end{table}

\begin{table}[!htbp]
\centering
\begin{tabular}{lp{0.75\textwidth}}
\toprule
Setting & High SNR $\boldsymbol{\beta}$ vector \\ \midrule
Even signal, $p=4$   & $\boldsymbol{\beta}_{9} = 0.66 \times (5,\; 4,\; -2.5,\; -3.7)$ \\[1ex]
Even signal, $p=10$  & $\boldsymbol{\beta}_{10} = 0.515 \times (5,\; 4,\; -2.5,\; -2.3,\; -3,\; 2,\; 0.5,\; 1.5,\; -3,\; 3,\; 2.5)$ \\[1ex]
Concentrated signal, $p=4$  & $\boldsymbol{\beta}_{11} = (4.5,\; 0,\; 0,\; 0)$ \\[1ex]
Concentrated signal, $p=10$ & $\boldsymbol{\beta}_{12} = (4.4,\; 0,\; 0,\; 0,\; 0,\; 0,\; 1.7,\; 0,\; 0,\; 0,\; 0)$ \\ \bottomrule
\end{tabular}
\caption{High SNR coefficient vectors for the \textbf{mixed covariate} setting. The medium and low SNR coefficients are defined as $\boldsymbol{\beta}_{9, \text{med}} = 0.575\,\boldsymbol{\beta}_{9}$ and $\boldsymbol{\beta}_{10,\text{low}} = 0.33\,\boldsymbol{\beta}_{10}$,
$\boldsymbol{\beta}_{11, \text{med}} = 0.575\,\boldsymbol{\beta}_{11}$, and $\boldsymbol{\beta}_{12, \text{low}} = 0.33\,\boldsymbol{\beta}_{12}$.}
\label{tab:beta_catcont}
\end{table}

\subsection{Categorical covariate distributions}
\label{app:a.2}

The categorical predictors were simulated using the distributions listed in Table~\ref{tab:cat_covs}. These distributions, in conjunction with the coefficient vectors described above, fully describe the data generation process used in the simulation study.

\begin{table}[h]
\begin{tabular}{ll}
\toprule
\textbf{Covariate} & \textbf{Distribution} \\ \midrule
\(X_1\)  & \(\operatorname{Bernoulli}(0.5)\) \\
\(X_2\)  & \(\operatorname{Bernoulli}(0.4)\) \\
\(X_3\)  & \(\operatorname{Bernoulli}(0.7)\) \\
\(X_4\)  & \(\operatorname{Bernoulli}(0.7)\) \\
\(X_5\)  & \(\operatorname{Multinomial}\) with \(P(0)=0.2,\;P(1)=0.25,\;P(2)=0.55\) \\
\(X_6\)  & \(\operatorname{Multinomial}\) with \(P(0)=0.2,\;P(1)=0.35,\;P(2)=0.45\) \\
\(X_7\)  & \(\operatorname{Bernoulli}(0.4)\) \\
\(X_8\)  & \(\operatorname{Multinomial}\) with \(P(0)=0.2,\;P(1)=0.35,\;P(2)=0.45\) \\
\(X_9\)  & \(\operatorname{Bernoulli}(0.45)\) \\
\(X_{10}\)  & \(\operatorname{Bernoulli}(0.55)\) \\ \bottomrule
\end{tabular}
\caption{Distributions for the categorical predictors; \(X_5,\ldots,X_{10}\) are included in the linear predictor only when $p=10$.}
\label{tab:cat_covs}
\end{table}

\section{FLC legend}
\label{app:b}

Table~\ref{table:flclegend} lists the factor levels associated with each FLC number. The first, second, and third sets of columns respectively denote FLCs that have all categorical covariates, all continuous covariates, and an equal number of categorical and continuous covariates.

\begin{table}[ht]
\centering
\scriptsize
\resizebox{\textwidth}{!}{%
\begin{tabular}{@{}ccccc|ccccc|ccccc@{}}
  FLC \# & \textbf{$p$} & \textbf{$n$} & SNR & Signal Dist.
  & FLC \# & \textbf{$p$} & \textbf{$n$} & SNR & Signal Dist.
  & FLC \# & \textbf{$p$} & \textbf{$n$} & SNR & Signal Dist. \\
  \cline{1-5}\cline{6-10}\cline{11-15}
  1   &  4 &  300 & H & Even   &  37 &  4 &  300 & H & Even   &  73 &  4 &  300 & H & Even   \\
  2   &  4 & 1200 & H & Even   &  38 &  4 & 1200 & H & Even   &  74 &  4 & 1200 & H & Even   \\
  3   &  4 & 2500 & H & Even   &  39 &  4 & 2500 & H & Even   &  75 &  4 & 2500 & H & Even   \\
  4   & 10 &  300 & H & Even   &  40 & 10 &  300  & H & Even   &  76 & 10 &  300  & H & Even   \\
  5   & 10 & 1200 & H & Even   &  41 & 10 & 1200 & H & Even   &  77 & 10 & 1200 & H & Even   \\
  6   & 10 & 2500 & H & Even   &  42 & 10 & 2500 & H & Even   &  78 & 10 & 2500 & H & Even   \\
  7   &  4 &  300 & M & Even   &  43 &  4 &  300  & M & Even   &  79 &  4 &  300  & M & Even   \\
  8   &  4 & 1200 & M & Even   &  44 &  4 & 1200 & M & Even   &  80 &  4 & 1200 & M & Even   \\
  9   &  4 & 2500 & M & Even   &  45 &  4 & 2500 & M & Even   &  81 &  4 & 2500 & M & Even   \\
 10   & 10 &  300 & M & Even   &  46 & 10 &  300  & M & Even   &  82 & 10 &  300  & M & Even   \\
 11   & 10 & 1200 & M & Even   &  47 & 10 & 1200 & M & Even   &  83 & 10 & 1200 & M & Even   \\
 12   & 10 & 2500 & M & Even   &  48 & 10 & 2500 & M & Even   &  84 & 10 & 2500 & M & Even   \\
 13   &  4 &  300 & L & Even   &  49 &  4 &  300  & L & Even   &  85 &  4 &  300  & L & Even   \\
 14   &  4 & 1200 & L & Even   &  50 &  4 & 1200 & L & Even   &  86 &  4 & 1200 & L & Even   \\
 15   &  4 & 2500 & L & Even   &  51 &  4 & 2500 & L & Even   &  87 &  4 & 2500 & L & Even   \\
 16   & 10 &  300 & L & Even   &  52 & 10 &  300  & L & Even   &  88 & 10 &  300  & L & Even   \\
 17   & 10 & 1200 & L & Even   &  53 & 10 & 1200 & L & Even   &  89 & 10 & 1200 & L & Even   \\
 18   & 10 & 2500 & L & Even   &  54 & 10 & 2500 & L & Even   &  90 & 10 & 2500 & L & Even   \\
 19   &  4 &  300 & H & Conc   &  55 &  4 &  300  & H & Conc   &  91 &  4 &  300  & H & Conc   \\
 20   &  4 & 1200 & H & Conc   &  56 &  4 & 1200 & H & Conc   &  92 &  4 & 1200 & H & Conc   \\
 21   &  4 & 2500 & H & Conc   &  57 &  4 & 2500 & H & Conc   &  93 &  4 & 2500 & H & Conc   \\
 22   & 10 &  300 & H & Conc   &  58 & 10 &  300  & H & Conc   &  94 & 10 &  300  & H & Conc   \\
 23   & 10 & 1200 & H & Conc   &  59 & 10 & 1200 & H & Conc   &  95 & 10 & 1200 & H & Conc   \\
 24   & 10 & 2500 & H & Conc   &  60 & 10 & 2500 & H & Conc   &  96 & 10 & 2500 & H & Conc   \\
 25   &  4 &  300 & M & Conc   &  61 &  4 &  300  & M & Conc   &  97 &  4 &  300  & M & Conc   \\
 26   &  4 & 1200 & M & Conc   &  62 &  4 & 1200 & M & Conc   &  98 &  4 & 1200 & M & Conc   \\
 27   &  4 & 2500 & M & Conc   &  63 &  4 & 2500 & M & Conc   &  99 &  4 & 2500 & M & Conc   \\
 28   & 10 &  300 & M & Conc   &  64 & 10 &  300  & M & Conc   & 100 & 10 &  300  & M & Conc   \\
 29   & 10 & 1200 & M & Conc   &  65 & 10 & 1200 & M & Conc   & 101 & 10 & 1200 & M & Conc   \\
 30   & 10 & 2500 & M & Conc   &  66 & 10 & 2500 & M & Conc   & 102 & 10 & 2500 & M & Conc   \\
 31   &  4 &  300 & L & Conc   &  67 &  4 &  300  & L & Conc   & 103 &  4 &  300  & L & Conc   \\
 32   &  4 & 1200 & L & Conc   &  68 &  4 & 1200 & L & Conc   & 104 &  4 & 1200 & L & Conc   \\
 33   &  4 & 2500 & L & Conc   &  69 &  4 & 2500 & L & Conc   & 105 &  4 & 2500 & L & Conc   \\
 34   & 10 &  300 & L & Conc   &  70 & 10 &  300  & L & Conc   & 106 & 10 &  300  & L & Conc   \\
 35   & 10 & 1200 & L & Conc   &  71 & 10 & 1200 & L & Conc   & 107 & 10 & 1200 & L & Conc   \\
 36   & 10 & 2500 & L & Conc   &  72 & 10 & 2500 & L & Conc   & 108 & 10 & 2500 & L & Conc   \\
\end{tabular}%
}
\caption{FLC Legend}
\label{table:flclegend}
\end{table}

\section{Additional MSE results}
\label{app:c}

\begin{figure}[ht]
\centering
\includegraphics[width=1.0\textwidth]{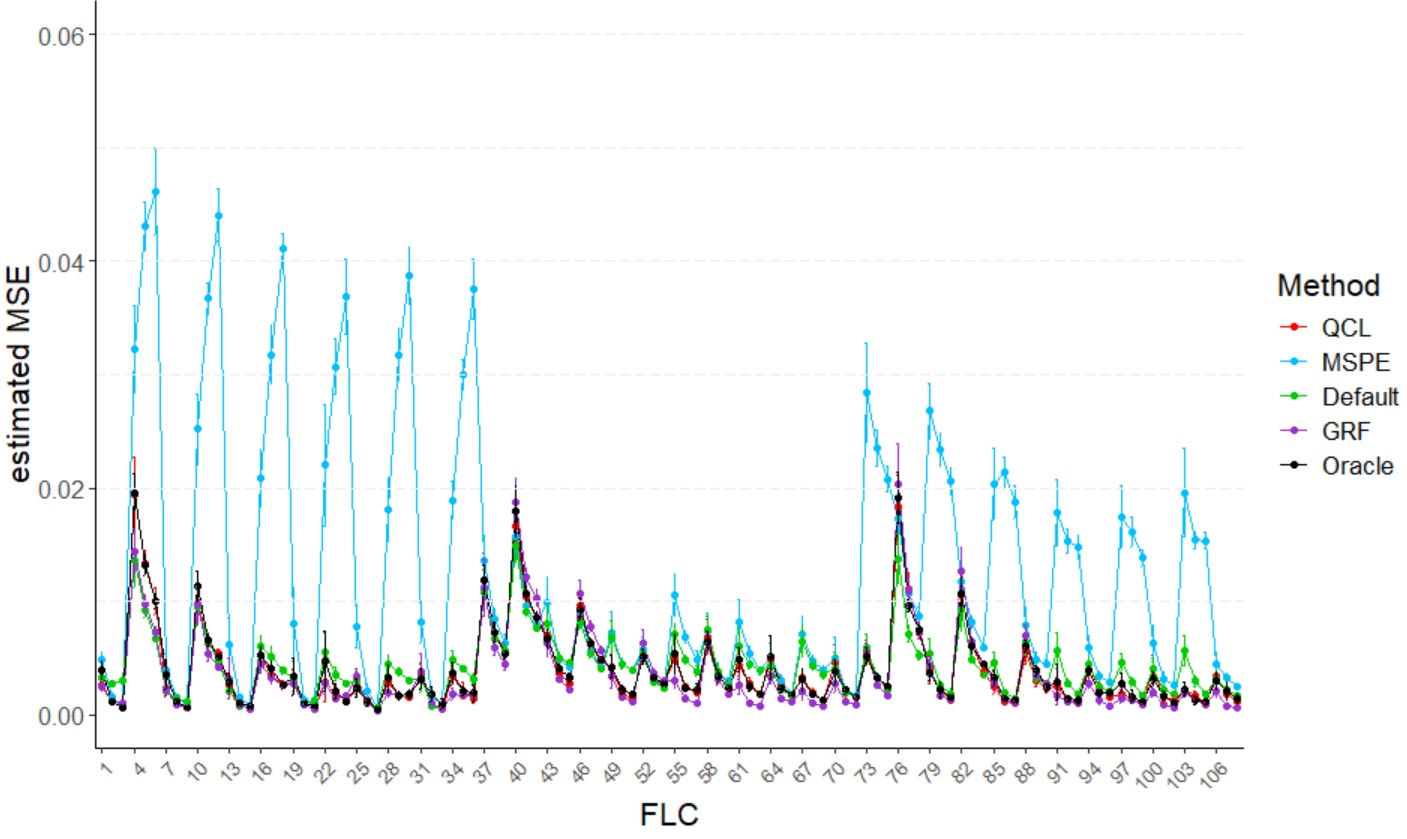}
\caption{Estimated MSE of coverage probabilities vs.\ FLC for each method ($\tau=0.1)$}
\label{fig_lineplots_covmse_uncensored}
\end{figure}

Figure~\ref{fig_lineplots_covmse_uncensored} shows estimated MSE of coverage probabilities vs.\ FLC for each method (similar to the analogous plot of coverage probability bias in Figure~\ref{fig_lineplots_covbias_uncensored}). The ideal is to achieve the lowest MSE in as many settings as possible. QCL tuning led to estimated MSE values that were, on average, comparable to (and often lower than) those produced using default tuning parameters and GRFs, whereas MSPE tuning clearly led to higher estimated MSE (in addition to higher estimated bias). It appears that QCL tuning does not result in excessively variable conditional coverage probabilities.

\section{Prediction interval coverage rates and widths}
\label{app:d}

\subsection{Summary statistics}

Tables~\ref{table_picovwidth_uncensored} and \ref{table_picovwidth_censored} display summary statistics (means, medians, first and third quartiles, minima, maxima, standard deviations) for the prediction interval coverage rates and widths in the uncensored settings (Table~\ref{table_picovwidth_uncensored}) and censored settings (Table~\ref{table_picovwidth_censored}).

\begin{table}[ht]
\centering
\resizebox{\textwidth}{!}{%
  \begin{tabular}{l ccccc ccccc}
    \toprule
      & \multicolumn{5}{c}{Coverage Rates} & \multicolumn{5}{c}{Widths} \\
    \cmidrule(lr){2-6} \cmidrule(lr){7-11}
               & Mean  & Median & Min   & Max   & SD  
               & Mean  & Median & Min   & Max    & SD   \\
    \midrule
    QRF-Default & 0.803 & 0.797  & 0.698 & 0.904 & 0.035 
                & 3.65  & 3.49   & 2.92  & 6.01   & 0.48  \\
    QRF-QCL     & 0.798 & 0.799  & 0.708 & 0.886 & 0.020 
                & 3.58  & 3.48   & 3.00  & 5.58   & 0.34  \\
    Res-OOB     & 0.800 & 0.800  & 0.729 & 0.872 & 0.017 
                & 3.51  & 3.46   & 2.92  & 4.82   & 0.22  \\
    Res-SC      & 0.799 & 0.800  & 0.684 & 0.889 & 0.023 
                & 3.56  & 3.50   & 2.67  & 5.75   & 0.29  \\
    \bottomrule
  \end{tabular}%
}
\caption{Coverage rates and widths of 80\% prediction intervals averaged across FLCs, uncensored data setting}
\label{table_picovwidth_uncensored}
\end{table}

\begin{table}[ht]
\centering
\resizebox{\textwidth}{!}{%
  \begin{tabular}{l ccccc ccccc}
    \toprule
      & \multicolumn{5}{c}{Coverage Rates} & \multicolumn{5}{c}{Widths} \\
    \cmidrule(lr){2-6} \cmidrule(lr){7-11}
                        & Mean  & Median & Min   & Max    & SD  
                        & Mean  & Median & Min   & Max    & SD  \\
    \midrule
    Default (10\%)      & 0.814 & 0.806  & 0.747 & 0.910  & 0.035 
                        & 3.74  & 1.28   & 0.76  & 36.43  & 7.08 \\
    QCL-C (10\%)        & 0.780 & 0.777  & 0.733 & 0.845  & 0.022 
                        & 3.46  & 1.24   & 0.75  & 35.42  & 6.78 \\
    QCL-IPCW (10\%)     & 0.803 & 0.804   & 0.759 & 0.845  & 0.015 
                        & 3.54  & 1.31   & 0.78  & 35.30  & 6.78 \\
    Default (30\%)      & 0.814 & 0.807  & 0.750 & 0.907  & 0.035 
                        & 3.89  & 1.27   & 0.76  & 38.05  & 7.26 \\
    QCL-C (30\%)        & 0.769 & 0.768  & 0.704 & 0.866  & 0.025 
                        & 3.42  & 1.20   & 0.74  & 36.00  & 6.76 \\
    QCL-IPCW (30\%)     & 0.829 & 0.828  & 0.742 & 0.903  & 0.031 
                        & 3.84  & 1.67   & 0.81  & 36.00  & 6.71 \\
    \bottomrule
  \end{tabular}%
}
\caption{Coverage rates and widths of 80\% prediction intervals averaged across FLCs, separated by censoring rate, for $n=1200$.}
\label{table_picovwidth_censored}
\end{table}

\subsection{Plots of prediction interval coverage rates}

\begin{figure}[ht]
\centering
\includegraphics[width=1.0\textwidth]{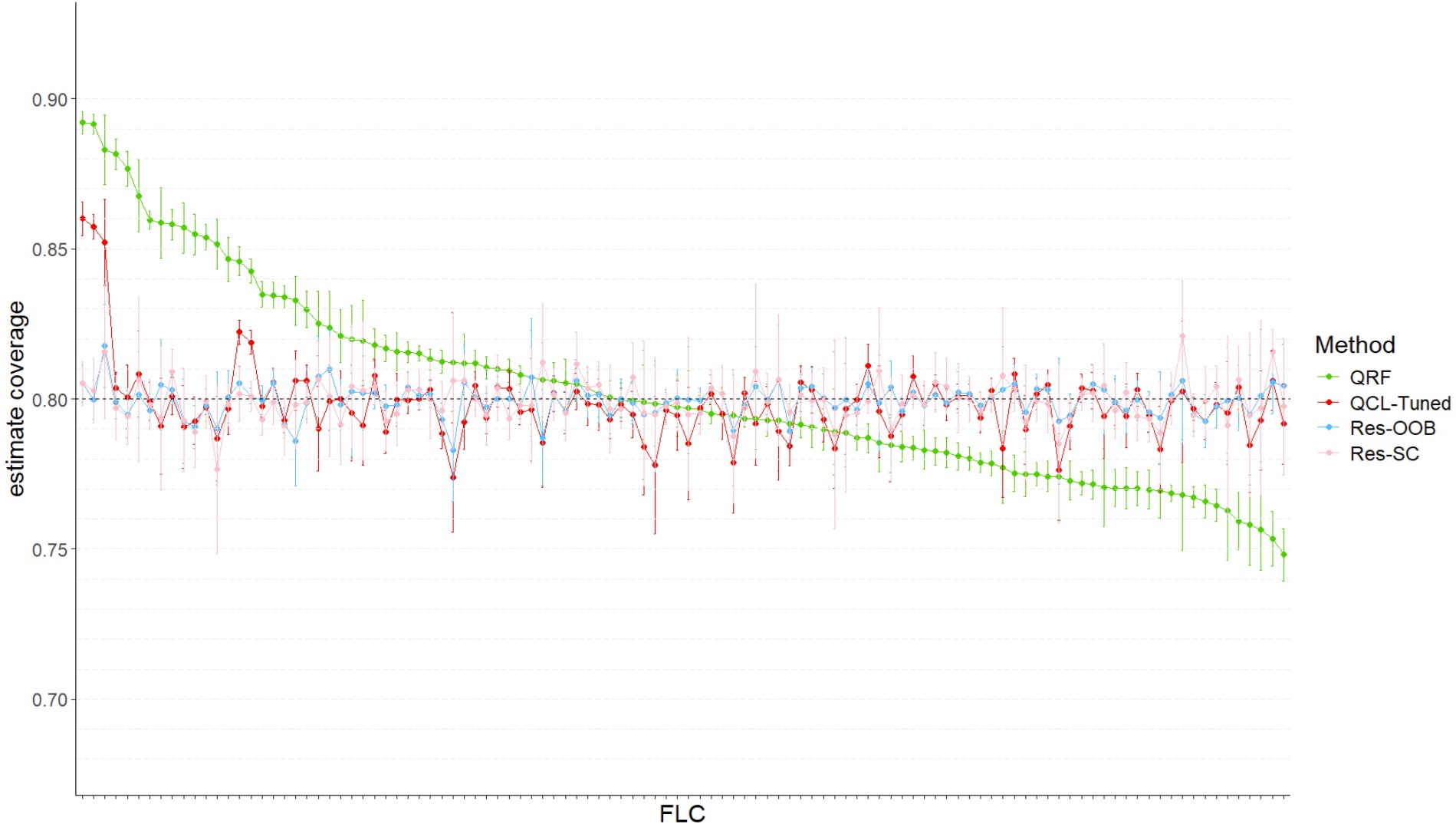}
\caption{Interval coverage rates vs.\ FLC in the uncensored data setting. The FLCs are ordered on the horizontal axis so that coverage rate according to QRF is descending.}
\label{fig_lineplots_picovbias_uncensored}
\end{figure}

\begin{figure}[ht]
\centering
\includegraphics[width=1.0\textwidth]{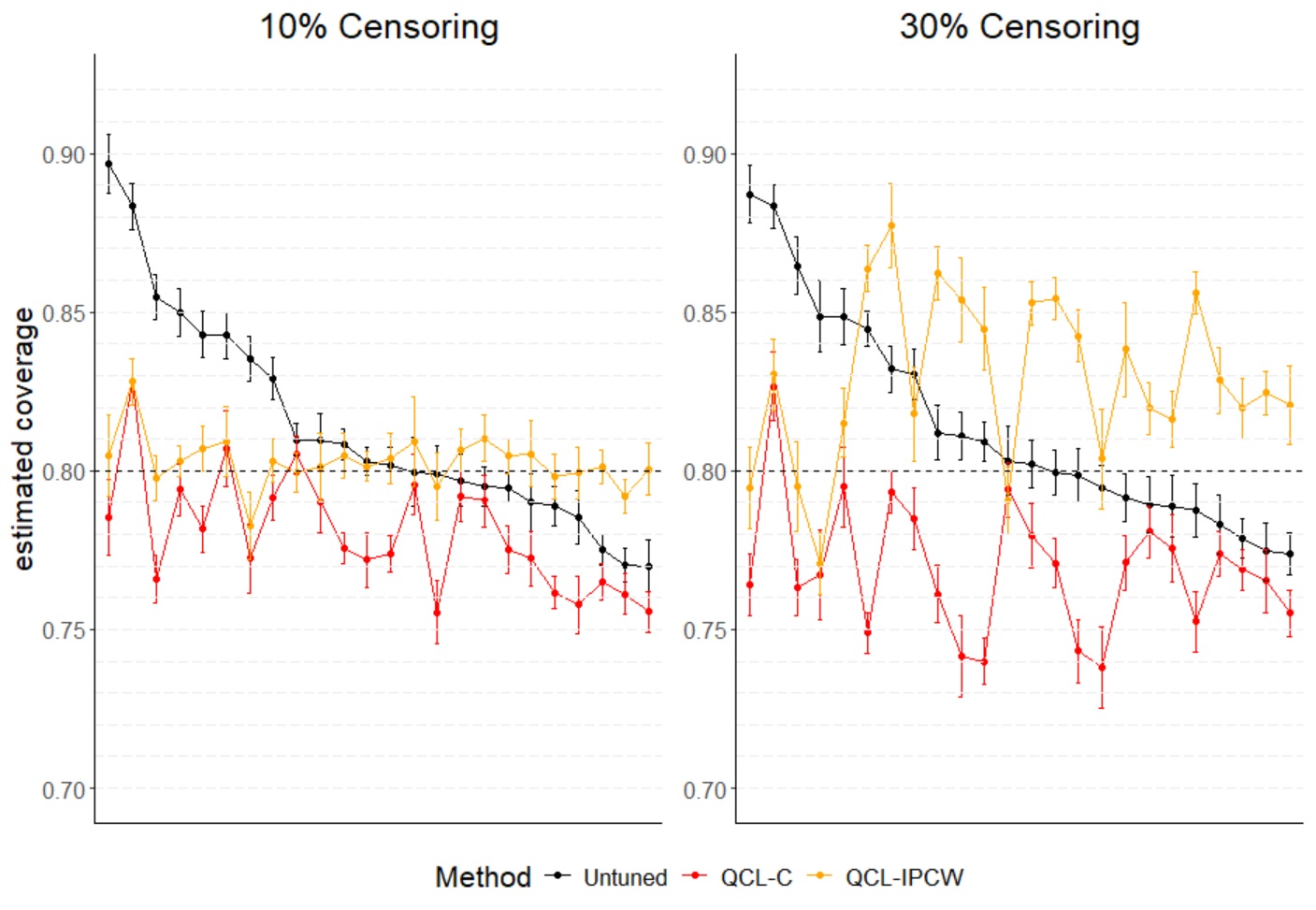}
\caption{Interval coverage rates by FLC in the censored data setting, broken down by censoring rate, for $n=1200$. The ordering of the FLCs on the horizontal axis is in descending order of coverage rate according to Default RSFs.}
\label{fig_lineplots_picovbias_censored}
\end{figure}

Figures~\ref{fig_lineplots_picovbias_uncensored} and~\ref{fig_lineplots_picovbias_censored} display plots of estimated prediction interval coverage rates (with confidence intervals) broken down by each FLC in the uncensored and censored data settings. The FLCs are ordered in terms of highest to lowest coverage probability of QRF intervals in the uncensored setting and Default RSF intervals in the censored setting.

\section{Estimated quantile bias}
\label{app:e}

\begin{figure}[ht]
\centering
\includegraphics[width=1.0\textwidth]{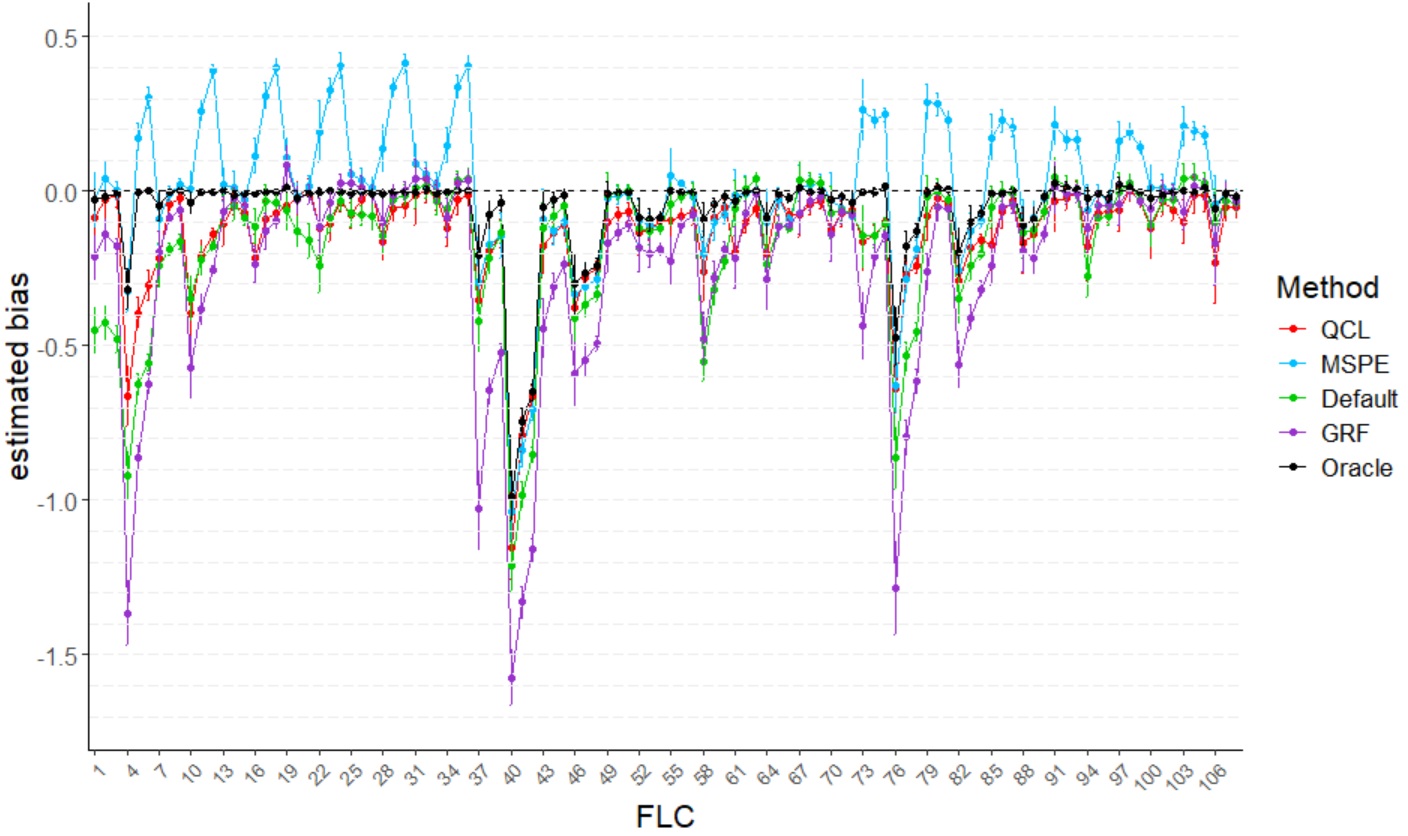}
\caption{Estimated quantile bias by FLC for each method ($\tau=0.1)$}
\label{fig_lineplots_qbias_uncensored}
\end{figure}

Figure~\ref{fig_lineplots_qbias_uncensored} displays a line plot of the estimated quantile bias for the 0.1 quantile broken down by each FLC in the uncensored data setting. Like in Figure~\ref{fig_lineplots_covbias_uncensored} in Section~\ref{sec5.1.1}, the sample means for each FLC are plotted for each method along with the Oracle, accompanied by t-based confidence intervals.

\section{The accuracy of estimates produced by QCL}
\label{app:f}

\begin{figure}[ht]
\centering
\includegraphics[width=1.0\textwidth]{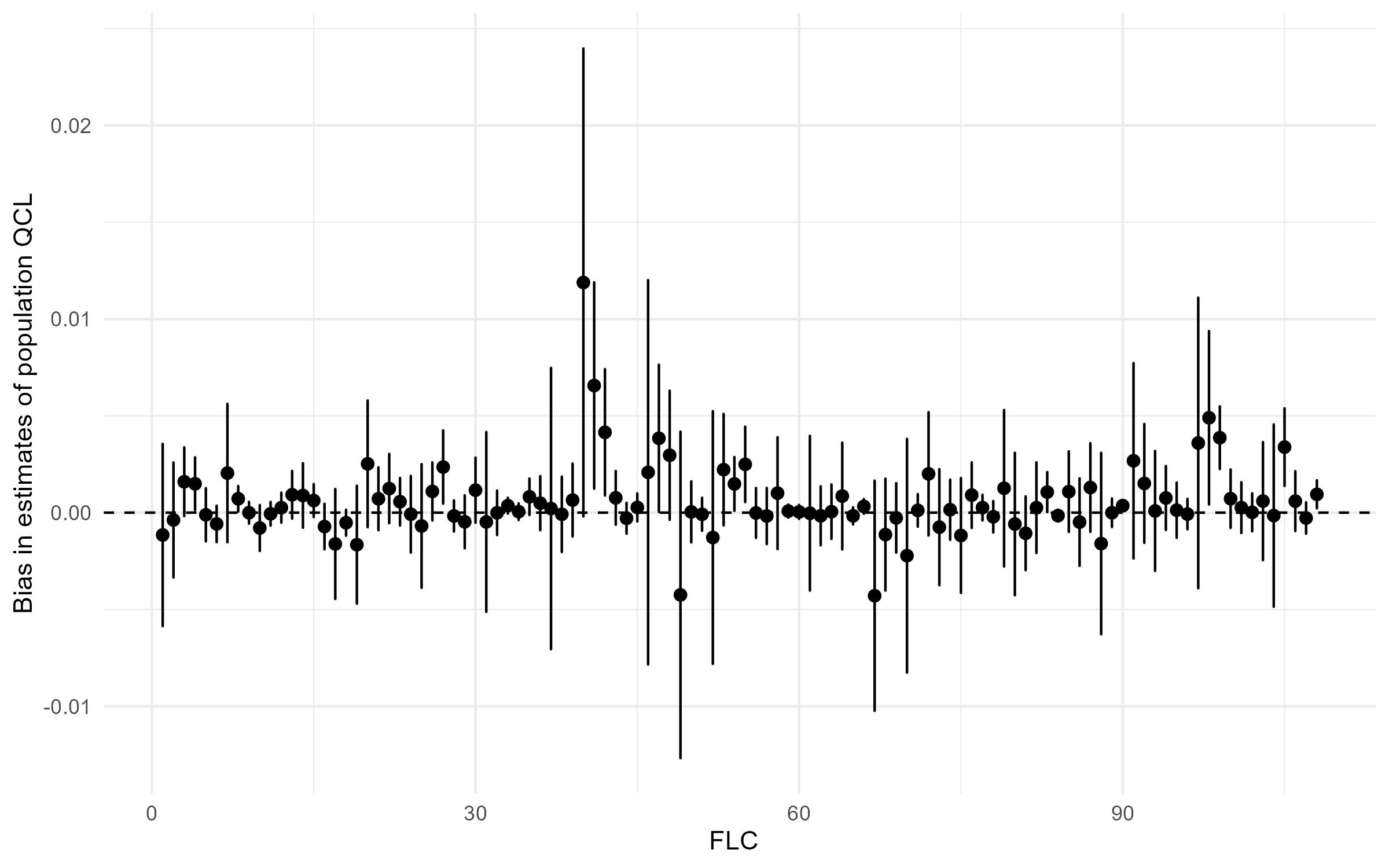}
\caption{Bias in estimates of the population QCL when estimating the $\tau=0.1$ quantile for the observations in the training set.}
\label{fig_qclvstruetrainingcoverage}
\end{figure}

In this appendix, we investigate the hypothesis that the estimates of the population QCL (based on the training set) in the uncensored-data setting were inaccurate (thus causing bias in our estimates of interest, i.e., coverage probabilities associated with observations in the test set). Figure~\ref{fig_qclvstruetrainingcoverage} displays the estimated bias in these estimates for each FLC in our uncensored data setting for $\tau=0.1$, along with confidence intervals. The estimated biases are computed as in Section~\ref{sec4.2} but using the training set instead of the test set.

The estimated biases are very close to 0 in nearly every setting. This finding is another piece of evidence that our tuning procedure is well-tailored to the estimation goal.

\noindent

\vskip 0.2in
\bibliography{qclpaper}

\end{document}